\documentclass[fleqn,usenatbib]{mnras}

\usepackage{newtxtext,newtxmath}

\usepackage[T1]{fontenc}
\usepackage{tablefootnote}

\DeclareRobustCommand{\VAN}[3]{#2}
\let\VANthebibliography\thebibliography
\def\thebibliography{\DeclareRobustCommand{\VAN}[3]{##3}\VANthebibliography}

\usepackage{graphicx}	
\usepackage{amsmath}	

\usepackage[normalem]{ulem}
\usepackage{xcolor}

\newcommand{\pks}{PKS\,0735+178}
\newcommand{\gr}{$\gamma$-ray}
\newcommand{\lsim}{{\lower.5ex\hbox{$\; \buildrel < \over \sim \;$}}}
\newcommand{\gsim}{{\lower.5ex\hbox{$\; \buildrel > \over \sim \;$}}}
\newcommand{\nup}{$\nu_{\rm peak}$}
\newcommand{\nufnu}{$\nu$f$({\nu})$}
\newcommand{\ic}{IceCube-211208A}
\newcommand{\ergs}{erg cm$^{-2}$s$^{-1}$}

\title[Neutrinos from PKS 0735+178]{A multi-messenger study of the blazar \pks{}: a new major neutrino source candidate }

\author[N. Sahakyan et al.]{
N. Sahakyan,$^{1,2, 3}$ \thanks{E-mail: narek@icra.it}
P. Giommi$^{4,5,6}$, 
P. Padovani$^{7,8}$,
M. Petropoulou$^{9}$\thanks{Mercator Fellow},
D. B\'egu\'e$^{10}$,
B. Boccardi$^{11}$,
S. Gasparyan$^{1}$\\
$^{1}$ICRANet-Armenia, Marshall Baghramian Avenue 24a, Yerevan 0019, Armenia\\
$^{2}$ICRANet, P.zza della Repubblica 10, 65122 Pescara, Italy\\
$^{3}$ICRA, Dipartimento di Fisica, Sapienza Università di Roma, P.le Aldo Moro 5, 00185 Rome, Italy\\
$^{4}$Center for Astro, Particle and Planetary Physics (CAP3), New York University Abu Dhabi, PO Box 129188 Abu Dhabi, United Arab Emirates\\
$^{5}$Institute for Advanced Study, Technische Universit{\"a}t M{\"u}nchen, Lichtenbergstrasse 2a, D-85748 Garching bei M\"unchen, Germany\\
$^{6}$Associated to Italian Space Agency, ASI, via del Politecnico snc, 00133 Roma, Italy\\
$^{7}$European Southern Observatory, Karl-Schwarzschild-Str. 
2, D-85748 Garching bei M\"unchen, Germany\\
$^{8}$Associated to INAF - Osservatorio di Astrofisica e Scienza dello Spazio, Via Piero 
Gobetti 93/3, I-40129 Bologna, Italy\\
$^{9}$Department of Physics, National and Kapodistrian University of Athens, University Campus Zografos, GR 15783, Athens, Greece \\
$^{10}$Bar Ilan University, Ramat Gan, Israel\\
$^{11}$Max-Planck-Institut f\"ur Radioastronomie, Auf dem H\"ugel 69, 53121 Bonn, Germany
}

\date{Accepted XXX. Received YYY; in original form ZZZ}

\pubyear{2022}

\begin{document}
\label{firstpage}
\pagerange{\pageref{firstpage}--\pageref{lastpage}}
\maketitle

\begin{abstract}
The blazar \pks\ is possibly associated with multiple neutrino events observed by the IceCube, Baikal, Baksan, and KM3NeT neutrino telescopes while it was flaring in the \gr, X-ray, ultraviolet and optical bands. 
We present a 
detailed  study of this peculiar blazar to investigate the temporal and spectral changes in the multi-wavelength emission when the neutrino events were observed. The analysis of Swift-XRT snapshots reveal a flux variability of more than a factor 2 in about $5\times10^3$ seconds during the observation on December 17, 2021. In the \gr\ band, the source was in its historical highest flux level at the time of the arrival of the neutrinos. The observational comparison between \pks\ and other neutrino source candidates, such as TXS 0506+056, PKS 1424+240, and GB6 J1542+6129, shows that all these sources share similar spectral energy distributions, very high radio and \gr\ powers, and parsec scale jet properties. 
Moreover, we present strong supporting evidence for \pks\ to be, like all the others, a masquerading BL Lac.
We perform comprehensive modelling of the multiwavelength emission from \pks\ within one-zone lepto-hadronic models considering both internal and external photon fields and estimate the expected accompanying neutrino flux. 
The most optimistic scenario invokes a jet with luminosity close to the Eddington value and the interactions of $\sim$ PeV protons with an external UV photon field. This scenario predicts $\sim 0.067$ muon and antimuon neutrinos over the observed 3-week flare. Our results are consistent with the detection of one very-high-energy neutrino like IceCube-211208A.
\end{abstract}




\section{Introduction}\label{sec:intro}

The discovery of a flux of very high-energy (VHE; $>$ 100 GeV) neutrinos of  astrophysical origin by the IceCube South Pole observatory\footnote{https://icecube.wisc.edu/}  \citep{Aartsen2013,IceCube10year} and the first reliable association of IceCube neutrinos with a cosmic source, the blazars TXS\,0506+056 \citep[e.g.,][]{TXS0506Science2018,Dissecting}, paved the way for the beginning of (extra-galactic) neutrino astronomy.
Recent works reporting hints, at various level of significance, of several other possible associations between IceCube neutrinos and blazars have been published \citep[e.g.,][]{IceCube10year,Plavin2020,Giommidissecting,Abbasi2021} strengthening the connection between VHE neutrinos and blazars \citep[see][for a recent review]{GiommiPadovani2021}.  

Blazars, a rare type of powerful Active Galactic Nuclei \citep[AGN,][]{AGNReview} with a relativistic jet pointing at the Earth \citep{Urry1995}, are known to be efficient and powerful cosmic accelerators, and for this reason have long been considered potential sources of astrophysical neutrinos \citep[e.g.][]{Stecker1991,Mannheim1993,HalzeZas1997,MuraseStecker2022}.
Blazars are sub-classified depending on their optical spectrum and on their radio to X-ray Spectral Energy Distribution (SED): sources showing broad emission lines are called Flat-Spectrum Radio Quasars (FSRQs), while sources with featureless optical spectrum, or displaying very weak emission lines, are called BL Lacs \citep[e.g.][]{Falomo2014}; the SED classification originally divides blazars into low- (LBLs), intermediate- (IBLs) and high-energy (HBLs) peaked sources \citep{PadovaniGiommi1995} or LSP, ISP and HSP \citep{Abdo2010}. These definitions have recently been refined into LBLs and intermediate-high-energy-peaked objects (IHBLs) depending on whether the peak frequency of the radio to X-ray SED (\nup\,) is located below or above $\sim 10^{13.5}$ Hz \citep[][]{GiommiPadovani2021}.

The IHBL object \pks\, is one of the brightest BL Lac objects in the sky. With a flux density of 2.3 Jy at 1.4 GHz in the NRAO VLA Sky Survey 
\citep[NVSS;][]{Condon1998}, this source is the fifth 
radio brightest BL Lac in the Roma-BZCat catalogue, 5th edition \citep{Massaro2015}. In the early 1990's, the radio flux density of \pks\, rose to the level of $\sim$ 5 Jy at 4.8 and 8 GHz \citep{Britzen2010} placing it among the brightest blazars of all types. 
\pks\, is also a bright source in the high energy (HE; $>100$ MEV) \gr\ band, with an average flux that is ranked no. 19 among the nearly 1,500 IHBL blazars included in the Fermi 4LAC-DR2  \citep{4LAC-DR2} 
catalogue. 
The optical spectrum of \pks\, is completely featureless and for this reason a precise redshift has not been measured yet. By detecting a strong absorption feature a lower limit of z $\geq$ 0.424 was provided by
\cite{Carswell1974}
(see also \citealt{2000A&A...357...91F} and \citealt{Rector2001}). The tentative detection of the host galaxy \citep{2012A&A...547A...1N} also implies a limit of $z=0.45\pm0.06$.
A value of z\,$\sim$ 0.65 has been recently proposed by \cite{Falomo2021} assuming that this source is a member of a group of faint galaxies detected close to its position. 
Even assuming $z = 0.424$, the corresponding radio and \gr\, luminosity are among the largest known in this type of sources, i.e. $L_{\rm R}$\,$\sim$ 10$^{27}$ W Hz$^{-1}$, $L_{\gamma}$ $\sim$ 10$^{47}$ erg s$^{-1}$.\\
\indent In this paper we present a multi-messenger study of \pks, which, in early December 2021, was found to be in spatial coincidence with multiple neutrino events by the IceCube \citep{GCN3119}, Baikal \citep{Baikal}, Baksan \citep{ATel15143}, and KM3NeT \citep{ATel15290} neutrino telescopes while it was going through its largest flare ever observed in the optical, UV, soft X-ray and \gr\, bands. This remarkable combination of events and multi-wavelength coverage  makes \pks\, one of the best candidate neutrino sources discovered so far.\\
\indent The paper is structured as follows. Section \ref{MMdata} presents the available multi-messenger data  and the multiwavelength data analyzed in this study. The multiwavelength light curve and SED of \pks\ are discussed in Section \ref{sedlc}. In Section \ref{sec:comparison} a comparison between \pks\ and other candidate neutrino sources is presented. In Section \ref{theory} the origin of the multiwavelength emission is investigated within one-zone lepto-hadronic models, and the discussion and conclusion are given in Section \ref{discon}. Throughout the paper, the following cosmological constants are adopted: $\Omega_M$ = 0.286 and $H_0$ = 69.3 km s$^{-1}$.
\section{Multi-messenger data}\label{MMdata}

The detection of \ic, Baikal, and Baksan neutrinos triggered 
a number of multi-frequency observations that found the source in a flaring state in various energy bands. The announcement of these early results via several astronomical telegrams \citep{ATel15021,Atel15105,ATel15108,ATel15113,ATel15136,ATel15143,ATel15148} was followed by other observations resulting in the rapid accumulation of a rich multi-frequency data set.
In this section we consider all the data that is currently public as well as the results of the analysis of proprietary data that have been published so far.

\subsection{Neutrinos from \pks\,}

The position of \pks\, is slightly outside the $\sim 13$ square degree 90 percent localization error (statistics only) of \ic\,\citep{GCN3119}, a track-like event with estimated energy of 172 TeV, and within the larger (5.5 degree, 50 percent containment) error region of one cascade Baikal neutrino with estimated energy of 43 TeV, and a chance coincidence probability of 2.85 sigma, detected 3.95 hours after the IceCube event \citep{Baikal}. The source was also reported to be in the error region of a GeV neutrino detected 4 days before by the Baksan Underground Scintillation Telescope, with a random coincidence probability of $\sim 3$ sigma \citep{ATel15143}. A follow up analysis of KM3NeT undersea neutrino detectors \citep{ATel15290} revealed the detection, on Dec 15, 2021, of an additional neutrino with an estimated energy of $\sim 18$ TeV and a p-value of the association with \pks\, of 0.14.  

\begin{figure}
\includegraphics[width=0.47\textwidth]{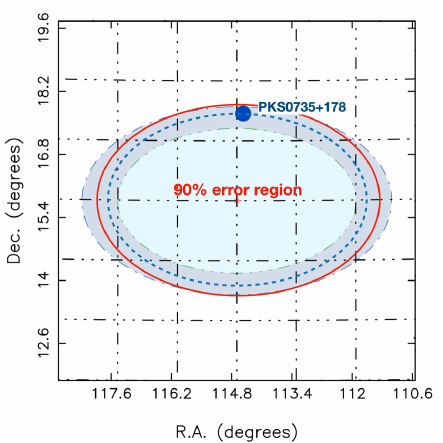}
	\caption{
	  The region around the localization area of \ic . The position of \pks\, is a few arcminutes north of the 90 percent error region (light blue area), but within the areas expanded to take into account possible IceCube systematics, according to the estimations of \citet[][darker blue area]{Giommidissecting}, \citet[][red line]{Plavin2020}, and \citet[][blue dashed line]{Hovatta2021}.
	}
\label{IC211208Error_region}
\end{figure}
  
Although \pks\, is located somewhat outside the 90 percent error region of \ic\, this does not preclude a real association since a modest offset is consistent both with the possible existence of a small IceCube systematic error \citep{Plavin2020,IceCubeSystematics,Hovatta2021} and with the obvious consideration that 10 percent of the real counterparts are expected to be outside the 90 percent error regions. In fact, the maximum signal for neutrino-blazar correlation in the work of \cite{Giommidissecting} was obtained by considering the 90 percent error region expanded by a factor 1.3. Following a different approach \cite{Plavin2020} estimated a systematic error of 0.5 degrees to be added linearly to the size of the IceCube error regions, while \citet{Hovatta2021} estimated that the systematic uncertainty is about 1.0 degree to be added in quadrature. Figure \ref{IC211208Error_region} shows that \pks\, is indeed inside in the error region of \ic\, when expanded to take into account systematic effects. 
A recent similar example is NGC1068, the AGN corresponding to the brightest excess in the 10 year IceCube neutrino sample \citep{IceCube10year}, which is also located outside the nominal 90 per cent error region of the track-like neutrino IceCube-211116A \citep{NGC1068GCN} but within the expanded area that takes into account systematic uncertainties, estimated according to the three methods mentioned above. 
\pks\,  might be the first source to be possibly associated to multiple neutrino events detected almost simultaneously by different telescopes.

\subsection{Fermi-LAT data}
The publicly available Fermi Large Area Telescope (LAT) data accumulated in the period from 2008 August 04 to 2022 February 15 have been analyzed using Fermi ScienceTools (1.2.1) and the  P8R3\_ SOURCE\_ V3 instrument response function. The 100 MeV–300 GeV PASS8 “Source” class events with a higher probability of being photons (evclass = 128, evtype=3) were  extracted from a $12^\circ$ region of interest (ROI) centered on the location of PKS 0735+178  [(R.A., decl.)=(114.54, 17.71)]. The events are binned within a $16.9^\circ \times 16.9^\circ$ square region into $0.1^\circ \times 0.1^\circ$ pixels and into 37 equal logarithmically spaced energy bins. The model file was created based on Fermi-LAT fourth source catalog (4FGL-DR2) where all sources within ROI+5 from
the position of PKS 0735+178 as well as the Galactic (gll\_ iem\_ v07) and the isotropic (iso\_ P8R3\_ SOURCE\_ V3) diffuse emission components are included. The binned likelihood analysis is applied to the entire data set using the {\it gtlike} tool.
The variation of gamma-ray flux was investigated by computing an adaptively binned light curve using the algorithm from \citet{2012A&A...544A...6L}. In this method, the overall period is divided into unequal time intervals with a constant uncertainty (20\% in this case) in each period. 

\subsection{Swift XRT and UVOT}
The Neil Geherels Swift Observatory \citep{SwiftPaper} observed \pks\, 27 times, 9 of which are after the arrival of \ic\,. 
We analysed all the  X-ray data from the X-Ray Telescope (XRT) using Swift\_xrtproc, a tool developed within the Open Universe initiative \citep{GiommiOU} that automatically performs a complete data reduction using the HEAsoftV6.29 software and generates high-level data products (assuming power law and log parabola spectral models and Galactic absorption), spectral and imaging analysis results, using the XSPEC\,V12.12.0 and XIMAGE\,V4.5.1 packages. Both data collected in single snapshots and over entire XRT observations\footnote{A snapshot is the time interval spent continuously observing a target. A Swift-XRT observation is composed of one or more snapshots.} are processed --
see \cite{Giommi2021} for details.
The results of our analysis are summarised in Table \ref{xrtresults} where column 1 gives the Modified Julian Day (MJD) of the observation, column 2 gives the power law (photon) spectral index, columns 3, gives the count rate, and columns 4 and 5 give the \nufnu\, flux at the energies of 1.0 and 4.5 keV respectively, demonstrating that most of the variability was confined to low energies.
Since the source was sufficiently bright to be detected in short exposures, for observations executed after \ic\, we also list the results of the analysis of each snapshot.

\pks\ was observed by the Swift Ultraviolet/Optical Telescope (UVOT)  
simultaneous with XRT. All the single observations were analyzed by the standard approach using HEASOFT v6.29. The source counts were extracted from a region of 5 arcsec radius centered at the source and the background counts were extracted from a region of 10 arcsec centered away from the source. The source magnitudes were extracted using {\it uvotsource} and were converted to fluxes using the conversion factors provided by \citet{2008MNRAS.383..627P}. Then, extinction corrections were applied using the reddening coefficient $E(B - V)$ from the Infrared Science Archive\footnote{http://irsa.ipac.caltech.edu/applications/DUST/}.

\begin{table}
\begin{footnotesize}
\begin{center}
\caption{Summary of Swift XRT observations of \pks\, after the detection of \ic.
}
\begin{tabular}{lcccc}
\hline\hline
MJD & Power law  & Count& \nufnu flux & \nufnu flux \\
 & index & $~~~$rate$^{(\rm a)}$ & $~~$1 keV$^{(\rm b)}$ & $~~$4.5 keV$^{(\rm b)}$\\
(1)&(2)&(3)&(4)&(5) \\
\hline
59558.38634$^{(\rm c)}$ &  2.8 $\pm$ 0.1 & 221.6 $\pm$  16. &  26.5 $\pm$   1.9 &   6.0 $\pm$   1.6 \\
59558.41761 &  2.6 $\pm$ 0.1 & 246.5 $\pm$  13. &  29.6 $\pm$   1.5 &  10.6 $\pm$   1.7 \\
59558.44985$^{(\rm c)}$ &  2.5 $\pm$ 0.1 & 282.9 $\pm$  22. &  33.5 $\pm$   2.4 &  17.9 $\pm$   3.6 \\
59560.40404 &  2.5 $\pm$ 0.1 & 170.4 $\pm$  15. &  18.3 $\pm$   1.6 &   8.9 $\pm$   2.2 \\
59561.57316$^{(\rm c)}$ &  2.6 $\pm$ 0.2 &  67.8 $\pm$   9. &   7.4 $\pm$   0.9 &   4.0 $\pm$   1.4 \\
59561.60169 &  2.5 $\pm$ 0.1 &  95.3 $\pm$   8. &   9.9 $\pm$   0.8 &   6.5 $\pm$   1.3 \\
59561.63469$^{(\rm c)}$ &  2.5 $\pm$ 0.2 & 130.5 $\pm$  14. &  13.6 $\pm$   1.5 &   9.7 $\pm$   2.4 \\
59562.40249 &  2.5 $\pm$ 0.2 & 112.2 $\pm$  12. &  13.3 $\pm$   1.3 &   3.8 $\pm$   1.3 \\
59562.43230$^{(\rm c)}$ &  2.5 $\pm$ 0.2 & 124.1 $\pm$  14. &  14.3 $\pm$   1.6 &   4.8 $\pm$   1.7 \\
59565.03193$^{(\rm c)}$ &  2.8 $\pm$ 0.1 & 293.5 $\pm$  20. &  32.4 $\pm$   2.2 &   7.5 $\pm$   1.9 \\
59565.09922$^{(\rm c)}$ &  2.7 $\pm$ 0.2 & 209.0 $\pm$  22. &  22.6 $\pm$   2.3 &   4.3 $\pm$   1.8 \\
59565.19582 &  2.7 $\pm$ 0.1 & 255.6 $\pm$  11. &  27.9 $\pm$   1.2 &   7.9 $\pm$   1.2 \\
59565.30356$^{(\rm c)}$ &  2.7 $\pm$ 0.2 & 372.8 $\pm$  34. &  41.7 $\pm$   3.7 &  11.5 $\pm$   3.5 \\
59565.36080$^{(\rm c)}$ &  2.5 $\pm$ 0.2 & 173.4 $\pm$  18. &  18.6 $\pm$   1.9 &   9.3 $\pm$   2.6 \\
59571.93520 &  2.3 $\pm$ 0.4 &  53.0 $\pm$  11. &   5.5 $\pm$   1.2 &   6.0 $\pm$   2.7 \\
59573.51501 &  1.8 $\pm$ 0.2 &  59.0 $\pm$   8. &   5.7 $\pm$   0.9 &   6.9 $\pm$   2.0 \\
59578.70018 &  1.7 $\pm$ 0.2 &  42.8 $\pm$   6. &   4.4 $\pm$   0.6 &   6.9 $\pm$   1.7 \\
59585.33808$^{(\rm c)}$ &  2.2 $\pm$ 0.4 &  27.4 $\pm$   7. &   3.7 $\pm$   0.9 &   2.5 $\pm$   1.5 \\
59585.40473 &  1.8 $\pm$ 0.2 &  33.2 $\pm$   5. &   4.0 $\pm$   0.6 &   5.2 $\pm$   1.5 \\
59585.47193$^{(\rm c)}$ &  1.6 $\pm$ 0.2 &  38.2 $\pm$   8. &   4.3 $\pm$   0.9 &   8.0 $\pm$   2.6 \\
\hline\hline
\end{tabular}
\label{xrtresults}
\end{center}
$^{(\rm a)}$ Units of counts/1,000 s, 0.3-10 keV;
$^{(\rm b)}$ Units of $10^{-13}$ \ergs\,\\
$^{(\rm c)}$ Single snapshot (red points in panels 2 and 3 of Fig. \ref{lcmulti}).

\end{footnotesize}
\end{table}

\subsection{Other X-ray data}
NuSTAR performed two target of opportunity 
observations of \pks, on December 11 and 13, 2021.
Preliminary results by \cite{ATel15113} reported an approximately constant flux of $\sim 3 \times 10^{-12}$ \ergs\, in the 3-40 keV band and a photon index of $\Gamma \sim$ 1.7, significantly harder than that seen by Swift-XRT. The difference between the XRT and NuSTAR spectral indices suggests that the latter detects the rising part of the high-energy component of the SED, while XRT probes the high-energy cutoff of the synchrotron spectrum. While the flux in the 3-40 keV band is almost constant,  most, if not all, of the X-ray variability occurred approximately below 4.5 keV, as it is evident from the Swift data reported in Table \ref{xrtresults}. Differences in flux variability between softer and harder X-rays are expected if these are produced by particles of different energies and/or the maximum energy of the particle distribution (or the acceleration timescale) is varying with time \citep[e.g.][]{2008A&A...491L..37M, 2014A&A...571A..83P}.

Historically, \pks\, was detected as an X-ray source multiple times with the Einstein IPC in 1979, 1980 and in 1981 \citep{Madejski1988}, with the EXOSAT CMA in 1983 \citep{Giommi1990}, and by the ROSAT satellite both during the RASS survey in 1990 \citep{RASS}, and in a pointed observation in 1992 \citep{WGA}. More recently \pks\, has been detected four times by SRG/ eROSITA \citep{2021A&A...647A...1P} during the on-going X-ray sky surveys  \citep{ATel15108}.
In all cases the estimated flux, converted to 1 keV \nufnu\, units was lower than or about $5\times 10^{-13}$ \ergs\,.
\begin{figure*}
\includegraphics[width=0.99\textwidth]{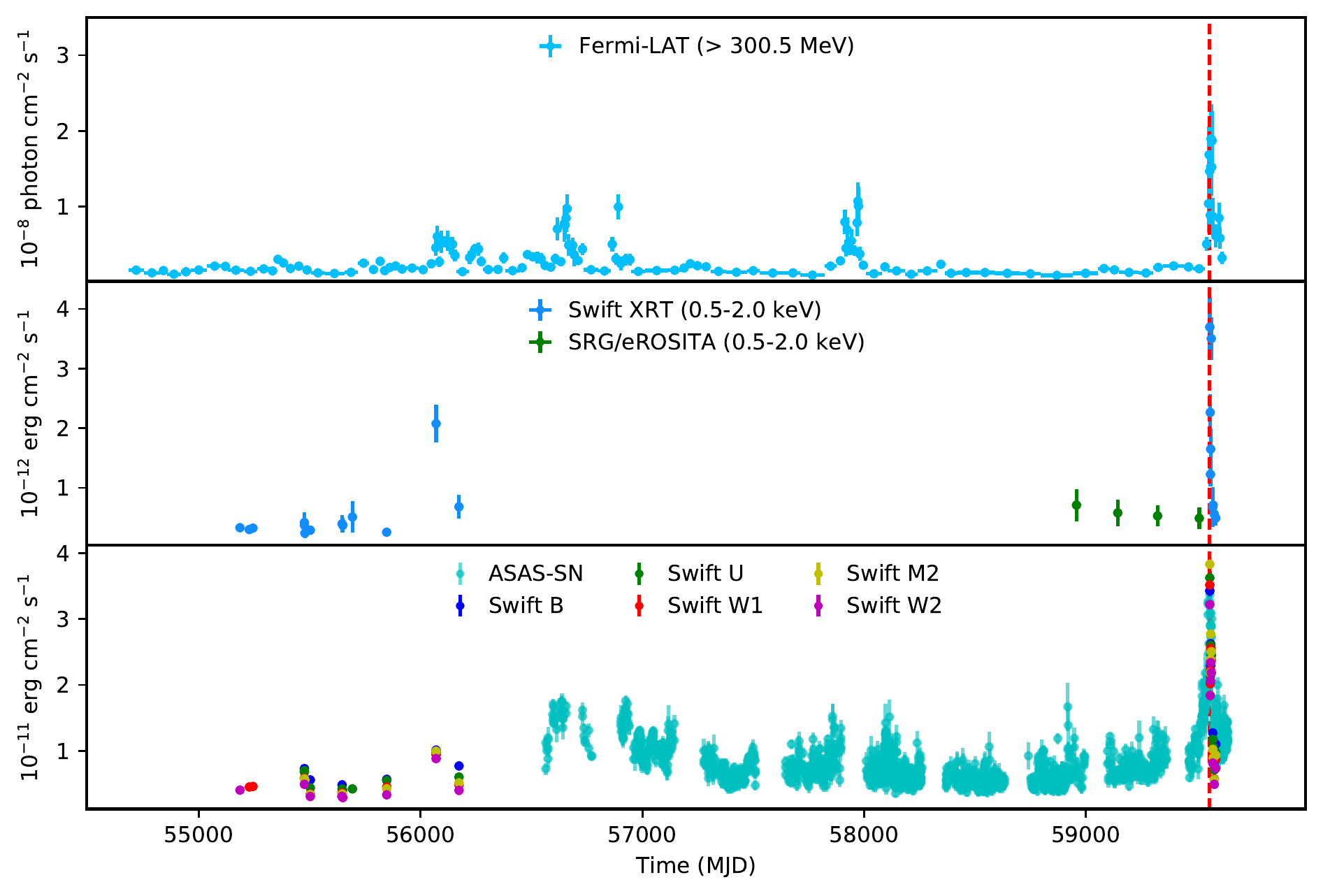}\\
\caption{Top panel: Fermi-LAT long-term  \gr\, lightcurve of \pks\, built with the adaptive-binning method. 
Middle and lower panels: X-ray and optical/UV lightcurve from ASAS-SN and Swift-UVOT data.
The vertical line marks the time of the arrival of \ic\,.
}
\label{lc}
\end{figure*}
\begin{figure}
\includegraphics[width=0.49\textwidth]{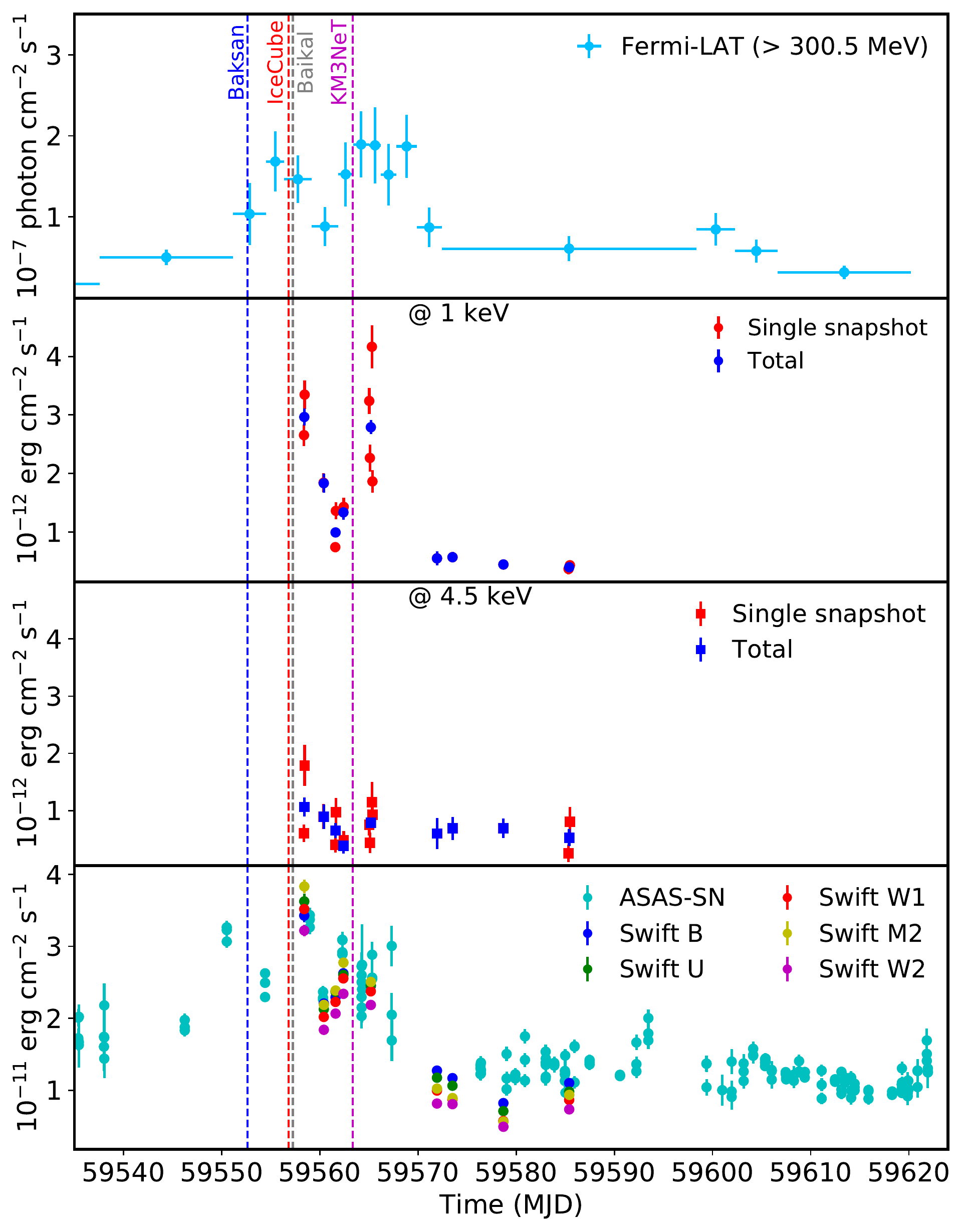}
	\caption{
	Multi-frequency light-curves near the time of detection of \ic\,. 
	 \label{lcmulti}}
\end{figure}

\begin{figure*}
\includegraphics[width=0.95\textwidth]{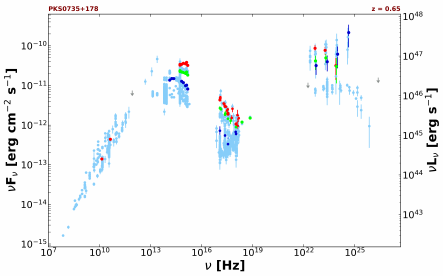}
	\caption{The SED of \pks\,. Light blue points are archival data. Red and green points refer to the time of the first and the second Swift observation after \ic\,. Dark blue points represent data collected at the end of the flare. Gray arrows represent upper limits. 
	\label{sed}}
\end{figure*}

\subsection{Other observations}
\pks\, is included in historical catalogues of  infrared (IRAS) and \gr\, (EGRET) sources. IRAS fluxes are much higher than other archival infrared data, while the  
EGRET flux is similar to that observed during the flare simultaneous with the neutrinos arrival.

In the optical band \pks\, was reported to be active since November 2021 when it was detected with an R magnitude of 14.88 and 8.5 per cent linear polarisation \citep{ATel15021}. 
This level of polarisation is not particularly large among BL Lacs, as the optical light in LBLs can be 40 percent polarised \citep{ImpeyTapia1990}. It is however close the highest level ($\approx$ 10 percent) observed in X-ray selected BL Lacs, which are typically blazars of the IHBL type \citep{Jannuzi1994}.
\pks\, has also been observed at near infrared frequencies after the IceCube announcement. The results of a number of observations, showing a decreasing flux in the days immediately after \ic\, has been reported by \cite{ATel15136} and \cite{ATel15148}.
In the radio band the source showed a slow but constant brightening months before the neutrino detections, nearly doubling its 37 GHz flux density from 0.6 Jy in January 2021 to 1.1 Jy at the time of \ic\,. The 
observations on 08 December 2021 show that the flux density between 14 and 44 GHz was $\sim$ ~1 Jy \citep{Atel15105}. 

\section{Multi-frequency lightcurve and SED of \pks}\label{sedlc}

The multi-wavelength light curve of \pks\ from 2008 to 2022 is shown in Fig. \ref{lc}. The Fermi-LAT \gr\, light curve of \pks , constructed with photons of energy larger than the optimal value of $300.5$ MeV using the adaptive-binning method of \citet{2012A&A...544A...6L}, is shown in Fig. \ref{lc} (top panel). The soft X-ray lightcurve constructed from the analysis of Swift-XRT data, and the optical/UV lightcurve assembled using ASAS-SN \citep[downloaded from the ASAS-SN
Sky Patrol web site,][]{ASAS-SN} and Swift-UVOT data are shown in middle and lower panels of Fig. \ref{lc}. For a comparison the data from eROSITA observations of \pks\ scaled to 0.5-2.0 keV range are shown with green points in the middle panel of Fig. \ref{lc}. The flux in all  the bands considered shows a similar behaviour: the largest flare since the launch of the Fermi satellite in 2008 occurred at the time of 
\ic\,, which is marked 
on the figure by a red vertical dashed line. Fig. \ref{lcmulti} displays a composite multi-frequency lightcurve around the time of the neutrino arrival.
The top panel shows that the \gr\ flux started increasing about three to five days before and peaked in correspondence of the neutrino arrival. A second flux increase reached an even higher maximum, with $F(E >300.5~\rm MeV) = (1.89\pm0.41)\times10^{-7}\:{\rm photon\:cm^{-2}\:s^{-1}}$, approximately 8 days later (on MJD $59564.21\pm0.83$). About 10 days afterwards the flux approximately returned to the pre-flare level. A similar evolution is apparent in the 1 keV lightcurve from the Swift monitoring plotted in the second panel 
of Fig. \ref{lcmulti}. 
No significant  variability is instead present at 4.5 keV (third panel), where the flux remains almost constant over all the observations after the neutrino arrival at a level somewhat higher than the historical flux. In the optical/UV band the flux started to increase several days before \ic\ and was above $3.0\times10^{-11}\:{\rm erg\:cm^{-1}\:s^{-1}}$ around MJD 59558.42 (fourth panel of Fig. \ref{lcmulti}). The source remained active in the optical/UV band until MJD 59570 then  returning to the pre-flare level.

The observations in the X-ray band on December 17, 2021 (MJD = 59565.19) resulted in the detection of fast variability of the soft X-ray flux between the third and the fifth XRT snapshot (red points in Fig.\ref{lcmulti} second panel). From Tab. \ref{xrtresults}, which lists the 1 keV and 4.5 keV \nufnu\, flux in each snapshot performed after \ic\,, 
there is about a factor two change in the 1keV flux in about  5~ks.

The SED of \pks, assembled with archival multi-frequency data retrieved with the VOU-Blazar tool \citep{voublazars} (light blue points) and with the data collected during the December 2021 flare (other colours) is plotted in Fig. \ref{sed}.
The Swift-XRT ToO monitoring shows that at the time of \ic\, the source was flaring in the soft X-ray band with a variable, mostly steep, spectral slope. Full details about the evolution of the SED of \pks\, in time is given in the SED animation available at \href{https://youtu.be/X58NotOjppg}{\nolinkurl{youtu.be/ipGJhh}} which shows the changes of broadband emission components before, during and after \ic\ event. \\
\indent The simultaneous optical and UV measurements show that \nup\, during the flare was definitively larger than $\sim $ 10$^{15}$ Hz. Fitting the optical to X-ray data to a polynomial function gives log(\nup) = 15.17 during the flare (red points in Fig. \ref{sed}) and log(\nup) = 14.13 after the flare (dark blue points in Fig. \ref{sed}), a value, this last, that is close to typical \nup\, values observed in archival data.
This range of \nup\, values, the fairly large synchrotron peak flux (a few times $10^{-11}$\ergs\,), the close to Jansky-level radio flux density, and the average \gr\ flux of $\sim 10^{-11}$\ergs\, at 1 GeV, make the overall SED of \pks\, qualitatively similar to that of other likely IBL/HBL bright neutrino emitters, namely TXS\,0506+056, PKS\,1424+240 and GB6 J1542+6129 \citep[although for the last two sources no multi-frequency observations close to the neutrino arrival are available and the presence of similar peak shifts cannot be tested, see Fig. \ref{sedscomparison} of this paper and Fig. 4 of][]{GiommiPadovani2021}.
Sources with these characteristics only make about 10 percent of the entire blazar population (only 183, out of a total of 1,711 blazars with radio flux density larger than 200 mJy\footnote{200 mJy is an indicative intensity approximately equal to the 1.4GHz flux density of GB6 J1542+6129, the faintest of the sources considered in this comparison. It is also the intensity above which existing catalogs are reasonably complete and include a sufficient number of blazars of all types to allow an accurate estimate of their relative abundances.} listed in current catalogs are of the IHBL type) and are intrinsically very rare, as there are only 18 such objects with radio flux density in excess of 1 Jy in the entire sky.
This peculiarity, combined with the observational evidence that large peak frequency changes correlated to source intensity are frequent in IHBL blazars while are rarely observed in LBLs \citep{GiommiPadovani2021} and with the results of  \citet{Giommidissecting}, who found a 3.23 $\sigma$ excess of IHBLs, and no excess of LBLs, in a large sample of IceCube tracks, all point in the direction of IHBLs possibly being the only type of blazars related to neutrino emission. However, this is still only suggestive and more data is necessary to confirm the connection.
\section{A comparison between PKS\,0735+178, TXS\,0506+056 and other candidate neutrino sources}
\label{sec:comparison} 
\subsection{\pks\, as a masquerading BL Lac} 
\label{subsec:masq}
\begin{figure*}
\includegraphics[width=0.95\textwidth]{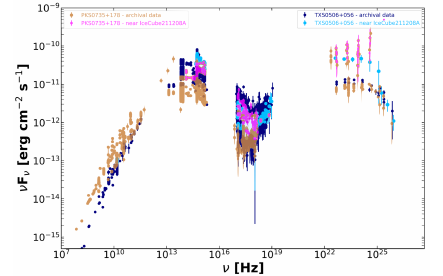}
	\caption{The SEDs of PKS\,0735+178 and of TXS\,0506+056 including archival data taken well before \ic\, (light brown and dark blue points), and data collected within two weeks after the neutrino detection (magenta and light blue points respectively). The two distributions are remarkably similar both in shape and intensity, especially shortly after the arrival of the neutrino when the optical to \gr\, data largely overlap.
	\label{sedscomparison}}
\end{figure*}

The evidence described above demonstrates that \pks\, can be considered one of the best candidate neutrino sources discovered so far. In this section we provide an observational comparison between \pks\, and TXS\,0506+056, the other blazar so far considered as the most likely example of an association between astrophysical neutrinos and a cosmic source.
Fig. \ref{sedscomparison} shows that the SEDs of the two sources are very similar, both in intensity and shape. The similarity is even more remarkable when considering the optical to \gr\, data collected during the two weeks following the arrival of the IceCube neutrinos (magenta and light blue points). In this period both sources show changing \nup\, values $\sim 10^{15}$ Hz, and highly variable X-ray and \gr\, fluxes.
\cite{Padovani2022b} showed that there are strong similarities between TXS\,0506+056 and other neutrino source candidates, such as PKS 1424+240 and GB6 J1542+6129, which have been found to be located in correspondence of neutrino excesses in the IceCube 10-year sample \citep{IceCube10year,Abbasi2021}. All these blazars not only possess nearly identical SEDs but also share other properties such as very high powers ($L_{\rm R} \gsim 10^{27}$ W Hz$^{-1}$, $L_{\gamma} \gsim 10^{47}$ erg s$^{-1}$), parsec scale properties (as estimated from very long baseline interferometry [VLBI] data: see Section \ref{subsec:VLBI}), and the unusual characteristics of being masquerading BL Lacs. 

\cite{Padovani2019} showed, in fact, that TXS\,0506+056 was, despite appearances, 
not a blazar of the BL Lac type but instead a masquerading BL Lac, namely
an FSRQ whose emission lines are swamped by a very bright, Doppler-boosted jet, 
unlike ``real'' BL Lacs, which are instead {\it intrinsically}
weak-lined. This is extremely relevant for two reasons: (1) ``real'' BL Lacs and FSRQs
belong to two very different physical classes, i.e., objects {\it without} and {\it
  with} high-excitation emission lines in their optical spectra, referred
to as low-excitation (LEGs) and high-excitation galaxies (HEGs),
respectively \cite[e.g.][and references therein]{AGNReview}; (2) 
masquerading BL Lacs, being HEGs, benefit from several radiation fields 
external to the jet (i.e., the accretion disc, photons reprocessed in the 
broad-line region (BLR) or from the dusty torus), which, by providing more 
targets for the protons, might enhance neutrino production as compared to LEGs. 
\cite{Padovani2022a} found a fraction of masquerading BL Lacs $> 24$ per cent 
in the \cite{Giommidissecting}'s sample. \cite{Padovani2022b} have shown that 
both PKS\,1424+240 and GB6\,J1542+6129, two IHBLs recently associated by IceCube 
with a neutrino excess, were also masquerading BL Lacs. One might then ask if \pks\, 
is also a masquerading source. \cite{Padovani2019} and 
\cite{Padovani2022a}, to which we refer the reader for more details, used
the following four parameters for this classification, in decreasing order of relevance: 
(1) location on the radio power 
-- \ion{O}{II} emission line power, $P_{\rm 1.4GHz}$ -- $L_{\rm [\ion{O}{II}]}$, 
diagram, which defines the locus of jetted (radio-loud) quasars; (2) a radio power 
$P_{\rm 1.4GHz} > 10^{26}$ W Hz$^{-1}$, since HEGs become the dominant population in the 
radio sky above this value; (3) an Eddington ratio\footnote{The Eddington luminosity is 
$L_{\rm Edd} = 1.26 \times 10^{46}~(M/10^8 \rm M_{\odot})$ erg s$^{-1}$, where $\rm M_{\odot}$ is 
one solar mass.} $L/L_{\rm Edd} \gtrsim 0.01$, which is typical of HEGs \citep[e.g.][and references therein]{Narayan1994,Fanidakis2011,AGNReview}; (4) a 
$\gamma$-ray Eddington ratio $L_{\gamma}/L_{\rm Edd} \gtrsim 0.1$. The latter two
parameters obviously require an estimate of the black hole mass, $M_{\rm BH}$.
We cannot use criteria (1) and (3) because the spectrum of \pks\, is featureless so we
have no handle on its $L_{\rm [\ion{O}{II}]}$ (needed for criterion (1)) and the derivation of the thermal, 
accretion-related bolometric luminosity  (needed for criterion (3)) requires an estimate of the emission line powers, which is not available\footnote{Note that an estimation of upper 
limits on emission line powers requires that redshift is available, as only then one knows 
where the line wavelength should be and can therefore derive the maximum flux value in order
for the line not be detected. However, by picking some redshift values 
$>0.424$ the resulting upper limits on $L_{\rm [\ion{O}{II}]}$ are consistent
with a masquerading BL Lac classification. For example, if we assume $z=0.65$ 
(Section \ref{sec:intro}) then $L_{\rm [\ion{O}{II}]} < 10^{41.6}$ erg s$^{-1}$, 
while for the corresponding $P_{\rm 1.4GHz}$ value of $10^{27.4}$ W Hz$^{-1}$ it 
should be $10^{41.2} < L_{\rm [\ion{O}{II}]} = 10^{42.2}$ erg s$^{-1}$.}.
However, based on its NVSS radio flux density (2.3 Jy) and a radio spectral index 
$\sim 0$ we derive $P_{\rm 1.4GHz} > 
10^{27}$ W Hz$^{-1}$ (since $z > 0.424$), i.e., well above the $10^{26}$ W Hz$^{-1}$ limit 
typical of HEGs (we stress that no LEGs can be as powerful as this). Although we do not
have a direct estimate of $M_{\rm BH}$ it is well known that blazar host galaxies are 
typical giant ellipticals \citep[e.g.][]{Padovani2022a}, which translates into 
$M_{\rm BH} \sim 10^{8.8\pm0.4} M_{\odot}$ (where we give the 1$\sigma$ dispersion: e.g. \citealt{Labita2006}). This implies $L_{\gamma}/L_{\rm Edd} > 10^{-0.1\pm0.4}$, 
due to the lower limit on redshift, that is well above the
0.1 limit for HEGs even taking into account the dispersion on $M_{\rm BH}$. In short, 
two out of four parameters are consistent with a masquerading BL Lac classification, 
while the remaining two cannot be used because we lack the relative information.  Therefore, following the analysis of \cite{Padovani2019,Padovani2022a}, we find that \pks\ should be classified as a masquerading BL Lac.
We further stress that its $L_{\gamma}$, $P_{\rm 1.4GHz}$, and \nup\ values put \pks\ into
a region of parameter space, which is {\it only} populated by masquerading BL Lacs (see Figs. 2 and 3 of \citealt{Padovani2022a}). 
Finally, we can also set a lower limit to the (hidden) BLR power by using the dividing line between ``real'' BL Lacs and FSRQs adopted by \cite{Ghisellini2011} of $L_{\rm BLR}/L_{\rm Edd} \sim 5 \times 10^{-4}$, which translates into $L_{\rm BLR} > 4 \times 10^{43}$ erg s$^{-1}$. 

\subsection{Radio properties of \pks}
\label{subsec:VLBI}
\cite{Padovani2022b} noticed some peculiarities in the radio band of the previously discussed 
neutrino candidates, which may be relevant for the production of neutrinos and may be shared also by \pks. 
These sources, in fact, being masquerading, are characterized by an accretion mode typical of powerful 
sources, and thus we would expect the production of jets 
with FSRQs-like radio properties. The jets in PKS\,1424+240 and TXS\,0506+056 were indeed found to be 
rather powerful based on their extended radio luminosity, which is within the Fanaroff \& Riley 
(FR: \citealt{Fanaroff1974}) II range ($\log P_{\rm ext}> 25.5$\,$\rm W Hz^{-1}$ at 1.4 GHz). 
In \pks, the extended radio power $P_{\rm ext}=7.5\times10^{24}-2\times10^{25}$\,$\rm W Hz^{-1}$, 
calculated assuming a spectral index $\alpha=0.8$ and the 1.4 GHz extended flux density reported 
by \cite{Rector2001}, approaches this same range. As in the other sources, however, this 
is not accompanied by the development of a clear FRII morphology. While noticing that the large scale 
morphology is 
difficult to define in blazars due to the strong projection effects, \cite{Padovani2022b} suggested that 
PKS\,1424+240 and 
TXS\,0506+056 may belong to the poorly populated class of FRI-HEG sources. As discussed by 
\cite{Perlman1994}, the large scale radio morphology of \pks\, resembles that of an FRI as well. 

A possible mismatch between the accretion mode and the radio properties was also shown to exist on VLBI 
scales. Indeed, the candidate neutrino sources are characterized by rather low apparent speeds $\beta_{\rm 
app} \equiv v_{\rm app}/c$ and core brightness temperatures $T_{\rm B}$, indicating modest values for the 
Doppler and Lorentz factors ($\delta_{\rm VLBI},\Gamma_{\rm VLBI}\lesssim 5$), as typically observed in HSP 
BL\,Lacs but not in FSRQs. Based on the results from the MOJAVE monitoring \citep{Lister2019}, the maximum 
jet 
proper motion observed in PKS\,0735+178 translates into an apparent speed $\beta_{\rm app}$ varying between 
6.7 and 9.7 for the adopted redshift range $z=0.424-0.65$, while the median core brightness temperature is 
$T_{\rm B}=2.6-3.2\times10^{11}$ K for the same range. Following the method adopted by \cite{Homan2021}, we 
can use the information on $\beta_{\rm app}$ and $T_{\rm B}$ to infer the Doppler and Lorentz factors, 
obtaining $\delta_{\rm VLBI}=6.8-7.9$ and $\Gamma_{\rm VLBI}=6.8-10.0$, again for the assumed redshift 
range\footnote{\cite{Homan2021} provide a lower limit on the median core brightness temperature of this 
source, $T_{\rm B}=2\times10^{11}$\,$\rm K$, calculated assuming $z=0$.}. When looking at the  ranges of 
maximum $\beta_{\rm app}$ ($\sim10 - 30$, e.g. \citealt{Jorstad2017}) and median core $T_{\rm B}$ 
($\sim10^{11}-10^{13}$\,$\rm K$, e.g. \citealt{Homan2021}) in $\gamma$-ray-detected FSRQs, these values 
lie at the lower end of such ranges in the case of \pks, and below them in the cases of PKS\,1424+240 and 
TXS\,0506+056. The same applies, then, to $\delta_{\rm VLBI}$ and $\Gamma_{\rm VLBI}$. 
The observed values are instead perfectly in line when considering, rather than the accretion mode, the 
spectral classifications of these jets as ISPs/HSPs, given the existence of a well-known anti-correlation 
between the maximum apparent speed in the jet and \nup\,
\citep{Lister2019}. \citet{Padovani2022b} speculated that the relatively rare combination of 
proton-loaded jets, possibly typical of high-excitation sources, and efficient particle acceleration 
processes, related to their relatively high \nup, might favour neutrino production in these sources (and not 
in FSRQs). We refer the reader to that paper for further details. 

\section{Theoretical modeling}\label{theory}
\begin{figure*}
\includegraphics[width=0.49\textwidth]{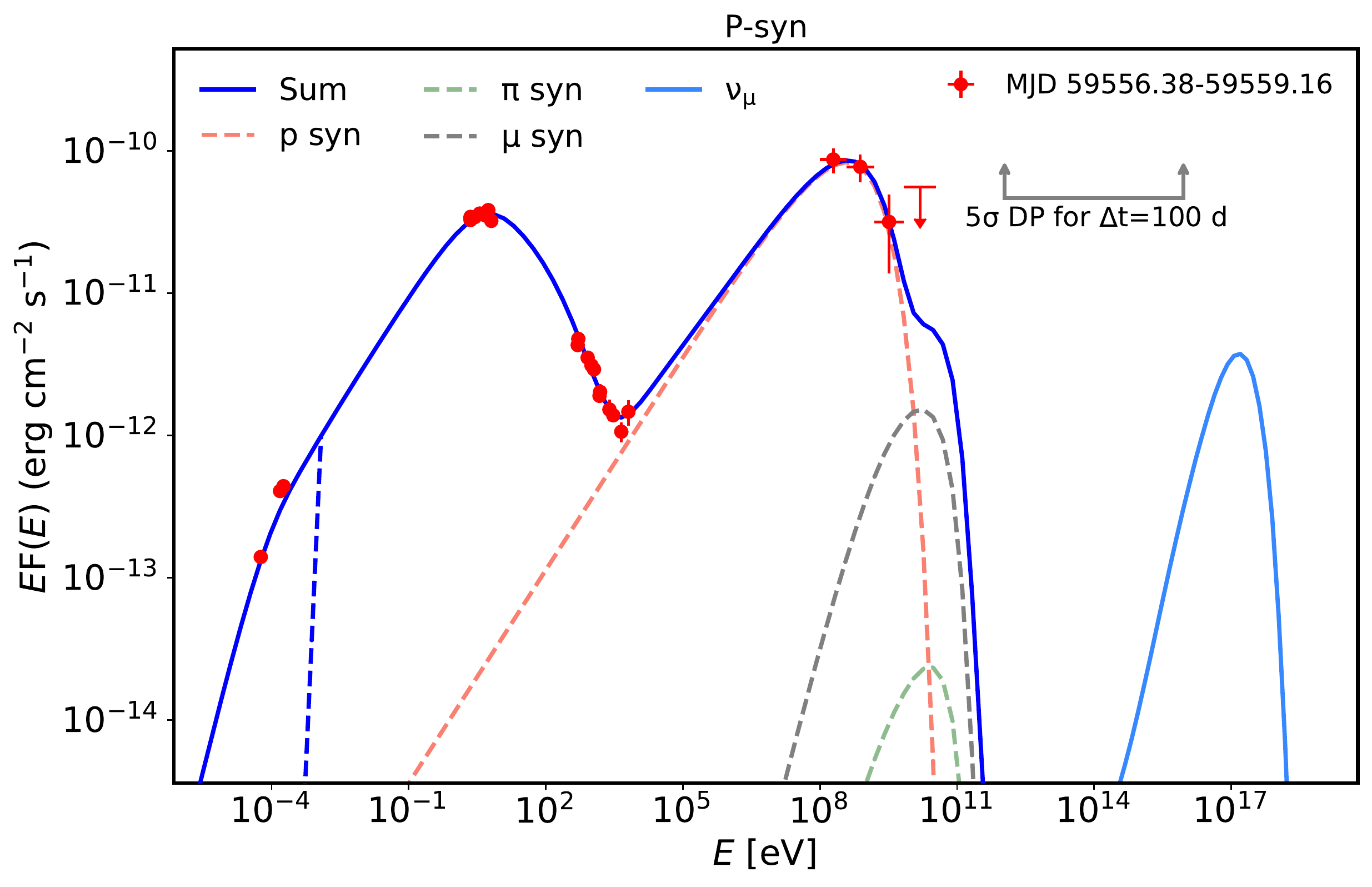}
\includegraphics[width=0.49\textwidth]{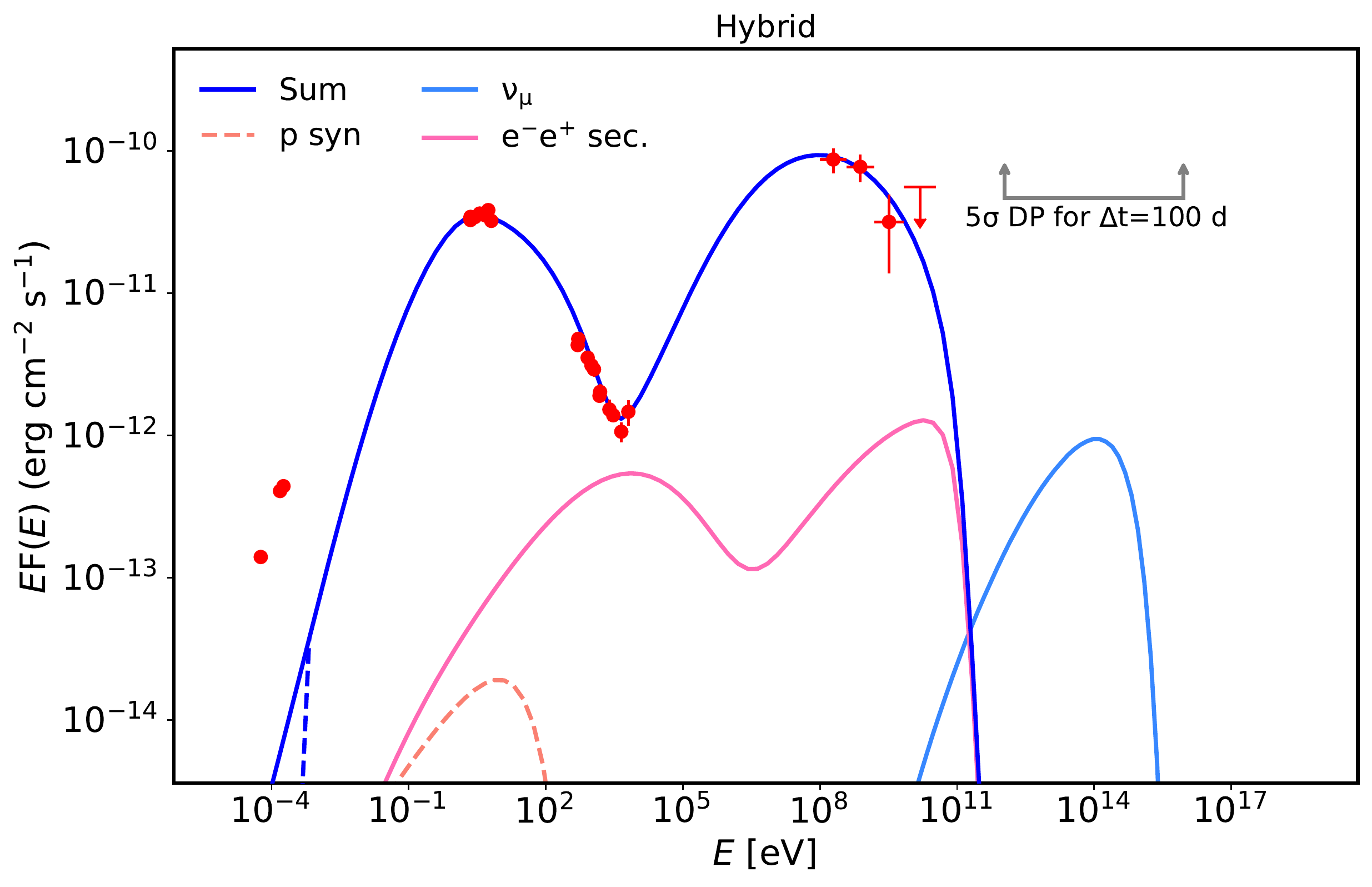}
	\caption{Broadband SED of \pks\ during the time of the arrival of IceCube-211208A modeled within P-syn (left panel) and hybrid (right panel) models. The solid blue line is the sum of all components taking into account \gr\ attenuation by EBL using the model of \citet{2011MNRAS.410.2556D} (for $z=0.65$). The blue dashed line shows the estimated spectrum with synchrotron self-absorption. The horizontal gray line corresponds to the $5\sigma$ (DP) for a flare duration of 100 days, assuming an $E^{-2}$ neutrino spectrum \citep[from][]{Abbasi2021}. For shorter duration flares the DP would move to the direction of the arrows. 
\label{sed_model}
}
\end{figure*}

Motivated by the similarities of the SEDs of \pks\, and TXS\,0506+056, illustrated in Fig.~\ref{sedscomparison}, we consider similar radiation models as those previously applied to the 2017 flare of TXS\,0506+056. More specifically,
we discuss three
scenarios: 
\begin{enumerate} 
\item a proton-synchrotron model (hereinafter P-syn) in which the high energy (hereinafter HE) component is mostly produced by proton synchrotron \citep[see e.g.,][]{2001APh....15..121M},
\item a hybrid model in which the low and high energy peaks are explained by leptonic processes and the maximum proton luminosity is constrained by the radiation in the X-ray band from the secondaries produced by the Bethe-Heitler and photo-pion processes \citep{2018ApJ...864...84K, 2019NatAs...3...88G, 2019MNRAS.483L..12C, 2022MNRAS.509.2102G}, and
\item  a hybrid model (hereinafter Hybrid-ext) where we also consider the presence of an external radiation field  as target for proton-photon interactions and inverse Compton scattering by relativistic leptons \citep[e.g.,][]{2018ApJ...864...84K, 2019ApJ...886...23X, Padovani2022a}.
\end{enumerate}

The code \textit{SOPRANO}\footnote{\url{https://www.amsdc.am/soprano/index.php}} \citep{2022MNRAS.509.2102G} is used to simulate the electromagnetic and neutrino emissions from \pks. The code has been developed to study the time-dependent \gr\ and neutrino emission from relativistic sources such as blazars and gamma-ray bursts, taking into account all relevant radiative processes but synchrotron self-absorption. In this paper, the spectra are produced under the steady state approximation. Considering a characteristic escape time equal to the dynamical time scale for all particles, we numerically compute the final spectrum by evolving the kinetic equations for several dynamical time scales to guarantee that the steady state is achieved. 

The emitting region is approximated by a sphere with a radius $R_{\rm}\leq \delta \:c\: t_{\rm var}/(1+z)$ inferred from the observed variability in the X-ray band $t_{\rm var}=5\times10^3$ sec (see Sec. \ref{sedlc}). The emitting region which moves with a bulk Lorentz factor $\Gamma\simeq\delta$, where $\delta$ is the Doppler factor, is seen in the direction close to the line of sight.
We assume that both electrons and protons are injected in the radiating region continuously. The distribution function of the electrons and protons at injection is assumed to be a power-law with exponential cutoff
\begin{equation}
Q^{\prime}_{\rm i}(\gamma_{\rm i})= \left \{ \begin{aligned}
& Q^{\prime}_{0,\rm i} \gamma^{-\alpha_{\rm i}}_{\rm i} \exp \left ( - \frac{\gamma_{\rm i}}{\gamma_{\rm i, cut}} \right ) & ~~~~ & \gamma_{\rm i, min} \leq \gamma_{\rm i}  \leq \gamma_{\rm i, max},  \label{dist} \\
& 0 & & {\rm otherwise,}
\end{aligned} \right.
\end{equation}
where $i = e, p$ for electrons and protons respectively. We assume that the proton and electron injection functions share a same spectral index $\alpha_{\rm e}=\alpha_{\rm p}$ and for the protons we assume $\gamma_{\rm p, cut}=\gamma_{\rm p, max}$.
The particles are injected in the radiation zone with a luminosity $L_{\rm i, jet}=\pi\: R^2\Gamma^2c U_{\rm i}$ ($i = e, p$) and $U_{\rm i}$ is the co-moving energy density of each particle, defined from their distribution function at injection as $U_{\rm i}=m_{\rm i}c^{2} \int \gamma_{\rm i}\:Q^{\prime}_{\rm i}(\gamma_{\rm i})d\gamma_{\rm i}$. Electrons and protons interact with a magnetic field of strength $B$ such that the magnetic luminosity is $L_{\rm B, jet}=\pi R^2 \delta^2c U_{\rm B}$. The electron synchrotron photons are target photons for the inverse Compton scattering, pair production, photo-pion and photo-pair production processes. VHE neutrinos are produced through the decay of charged pions, while energetic photons can be produced via $\pi^0$ decay and inverse Compton scattering. Usually, the optical depth for HE photons (in the emitting region) to photon-photon pair production is larger than unity, hence an electromagnetic cascade is triggered, transferring energy to lower energy photons.

\begin{figure}
\includegraphics[width=0.49\textwidth]{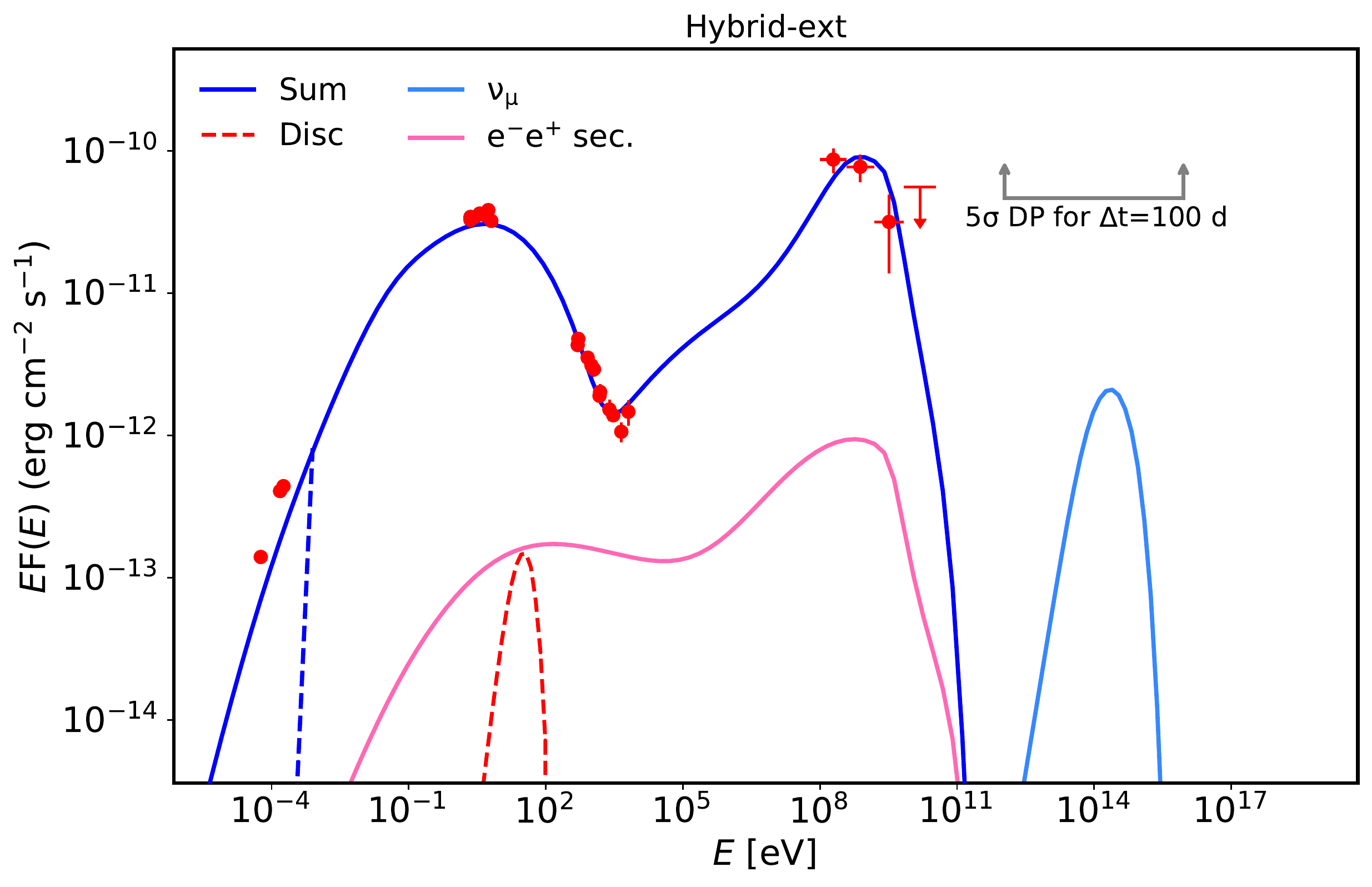}
	\caption{Same as in Fig. \ref{sed_model} but taking into account the BLR radiation field with a luminosity of $L_{\rm BLR} = 4 \times 10^{43}$ erg s$^{-1}$. The red dashed line is the black-body approximation to the disc emission.
\label{sed_hybrid_ext}
}
\end{figure} 
Since \pks\ is a masquerading BL Lac, we also consider the presence of the external radiation field provided by the BLR (see hybrid-ext scenario). We assume that the BLR is a spherical shell of a radius $R_{\rm BLR}=10^{17}\:L_{\rm BLR,44}^{0.5}$ cm \citep{2008MNRAS.387.1669G} and is characterized by a photon energy density $u_{\rm BLR}=L_{\rm BLR}/(4\pi R_{\rm BLR}^2 c)$. Here, $L_{\rm BLR}$ is the BLR luminosity  which is estimated to be $> 4 \times 10^{43}$~erg s$^{-1}$ (see Sec.~\ref{sec:comparison}). The comoving energy density is $\approx \Gamma^2 u_{\rm BLR}$ assuming that the emission region lies within the BLR \citep{1996MNRAS.280...67G}. The energy spectrum of the BLR radiation is modelled as 
a black body with a peak in $\nu F_{\nu}$ units at $2\times10^{15} \, \Gamma$ Hz and as measured in the comoving frame \citep{2008MNRAS.387.1669G}. The HE \gr\ data considered here from Fermi-LAT observations (up to tens of GeV) are below the threshold energy for  $\gamma-\gamma$ absorption, so these photons will escape the BLR.

Fig. \ref{sed_model} shows the results of our modelling for the P-syn and hybrid emission scenarios, whose parameters are given in Table \ref{tabel_param}. The blue solid line is the steady state photon spectrum considering all processes. Both our models can satisfactorily explain the data observed in optical/UV, X-ray and \gr \ bands. However when taking into account the synchrotron self-absorption process which produces $\nu F\nu\sim \nu^{7/2}$ spectrum at lower frequencies shown by dashed blue lines, the radio data cannot be explained. The emission in the radio band could be produced from low-energy electrons, which are perhaps located in more
extended jet regions.

The P-syn and hybrid models applied herein require very different magnetic fields and initial injection spectra for both protons and electrons. For the same size of the emitting region, $R=2.8\times10^{15}$ cm, and for the same Doppler boost, $\delta=30$ \footnote{This Doppler factor is larger than that estimated from VLBI data in Sec. \ref{sec:comparison}, which is related to the well-known and so-called “Doppler factor crisis” for HBLs \citep[e.g., see][]{2006ApJ...640..185H,2006tmgm.meet..512T}.
}, the magnetic field is $B=120$ G in the P-syn model, while it is significantly smaller for the hybrid model with $B=1.8$ G. This impacts the cut-off energy of the injected electrons: they should be accelerated up to $\gamma_{\rm e, cut}=1.4\times10^4$ ($\rm {E_{e,cut}}=7.2$ GeV) in the hybrid model to properly explain both components in the SED, as compared to $\gamma_{\rm e, cut}=1.9\times10^3$ for the P-syn model. Although in both models the protons are injected with the same power-law index, different maximum energies of protons are required. In the P-syn model the proton distribution should extend up to $\gamma_{\rm p,max}=3.0\times10^8 ~ ({\rm E_{p,max}}=2.81\times10^{17}\:{\rm eV})$ to explain the HE peak\footnote{The Larmor radius of protons with Lorentz factor $\gamma_{\rm p,max}=3.0\times10^8$ is smaller than the radius of the emitting region. Therefore provided that adequate conditions are met, protons might be accelerated to such a high energy.}. The proton synchrotron component is shown by the red line in Fig. \ref{sed_model} left panel. In this case, the muon and pion synchrotron emission, displayed by the grey and green dashed lines in Fig. \ref{sed_model} left panel,  contributes only in the VHE \gr\ band. This is similar to the model discussed for Mrk 421 in \citet{2015MNRAS.448..910C}.  
Instead, for the hybrid model, as the protons do not directly contribute to the observed SED, their maximum energy is smaller with $\gamma_{\rm p,max}=3.5\times10^5$. This is large enough to produce the bulk of the neutrino emission around the energy of IceCube-211208A event ($\sim$172 TeV). These protons are also interacting with jet photons via the Bethe-Heitler process, producing secondary energetic electrons which cool via synchrotron and inverse Compton processes producing a broad emission spectrum, see the magenta line in Fig. \ref{sed_model} (right panel). The amount of secondary emission is mostly constrained by the flux in the X-ray band, which in turns constrained the maximum proton content.   

\begin{table}
    \centering
    \caption{Parameter values used for the SED models in Figs. \ref{sed_model} and \ref{sed_hybrid_ext}.}
   		\begin{tabular}{@{}l c c c}
 		\hline
		 & P-syn & Hybrid & Hybrid-ext$^{(\rm a)}$ \\
 		\hline
 $\delta$ & $30$  & $30$ & $30$ \\
 $R~(10^{15}\:{\rm cm})$ & $2.8$ & $2.8$ & $2.8$ \\
 $B~({\rm G})$ & $120$  & $1.8$ & $5.9$ \\
 $\gamma_{\rm e, min} $& $300$  & $1.4\times10^3$ & $1.4\times10^2$ \\
 $\gamma_{\rm e, cut}$& $1.9\times10^3$  & $1.8\times10^4$ & $7\times10^3$ \\
 $\gamma_{\rm e, max}$& $2\times10^6$  & $5\times10^4$& $2.3\times10^4$\\
 $\alpha_{\rm e}$ & $2.0$  & $2.0$ & $1.9$ \\
 $\alpha_{\rm p}=\alpha_{\rm e}$ &  2.0 & 2.0 & $1.9$ \\
 $\gamma_{\rm p, min}$& 1 & 1 & 1\\
 $\gamma_{\rm p,max}$& $3.0\times10^8$  & $3.5\times10^5$ & $3.5\times10^5$ \\ \hline
 $L_{\rm e, jet}\:({\rm erg\:s^{-1}})$& $3.35\times10^{44}$  & $1.82\times10^{45}$ & $1.20\times10^{45}$\\
 $L_{\rm B, jet}\:({\rm erg\:s^{-1}})$ & $3.81\times10^{47}$  & $8.57\times10^{43}$& $9.20\times10^{44}$\\ 
 $L_{\rm p, jet}\:({\rm erg\:s^{-1}})$& $2.63\times10^{47}$ & $1.36\times10^{50}$ & $3.06\times10^{47}$\\
 		\hline
 		\end{tabular}
 	 \label{tabel_param}

$^{(\rm a)}$ The radiation from the BLR is modelled as a grey body with a peak energy at $2 \times 10^{15}$ Hz and a luminosity of $L_{\rm BLR} = 4 \times 10^{43}$ erg s$^{-1}$.
 \end{table}
The SED 
of the hybrid-ext model is shown in Fig. \ref{sed_hybrid_ext} (see Table \ref{tabel_param} for the parameters). 
In this case, the HE component  comprises of SSC radiation, which dominates the X-ray band, and inverse Compton scattered radiation of BLR photons dominating the $\gamma$-ray band.
Since the comoving temperature of the BLR is large, the electron distribution function does not need to extend to a large Lorentz factor. We find that the hybrid-ext model requires electrons to be accelerated only up to $\gamma_{\rm e,cut}=7\times10^3$, about two times smaller than the requirement of $\gamma_{\rm e,cut}=1.8\times10^4$ for the hybrid model. Instead, the maximum energy of the proton distribution is the same in both cases resulting in a similar peak energy for the predicted neutrino distribution. However, both the power-law index and luminosity are different. They are constrained from the X-ray band, considering the radiation from the secondaries to be sub-dominant compared to the SSC radiation of the primary electrons. Also, the spectrum  from secondaries has a different shape. 
In the hybrid-ext model, the low-energy peak from the synchrotron emission of the Bethe-Heitler pairs is lower than the HE peak (from photo-pion secondaries) while in the hybrid model, these two peaks have comparable peak flux. 
 		
Having estimated the luminosity of protons and their energy
distribution, the predicted neutrino spectra can be calculated straightforwardly. 
The muon neutrino spectra\footnote{Neutrino oscillations are taken into account assuming vacuum neutrino mixing and using 1/3 to convert the all-flavour to muon neutrino flux.} are shown by the light blue lines in Fig. \ref{sed_model} which are compared with the $5\sigma$ flare discovery potential (DP), assuming
a neutrino spectral index $\gamma_{\rm f}=2$ and a flare duration of 100 days \citep{Abbasi2021} \footnote{We note, however, a direct comparison cannot be made because the assumed spectrum for computing the discovery potential is different than the one in our models.}. For the considered models, the neutrino spectra peak at different energies: the P-syn model with a higher maximum energy of protons leads to a peak energy of the neutrino spectrum at $\sim10^{17}$ eV, much higher than that of the hybrid models at $\sim2\times10^{14}$ eV. These results are in agreement with previous studies of blazar spectra in the context of hadronic models \citep[e.g.][]{2014APh....54...61D, 2018ApJ...864...84K, 2021ApJ...912...54R}.

The expected number of muon and antimuon neutrinos for the considered models can be computed using the energy dependent point-source effective area of IceCube from \citet{IceCube10year}  for the declination of \pks. For $E_{\rm \nu, min}=100$ GeV and $E_{\rm \nu, max}=10^9$ GeV minimum and maximum energy of the neutrinos, the expected number of neutrinos during the $\sim$21.3 days flaring activity of \pks\ is $0.013$, $0.037$ and $0.067$ for the P-syn, hybrid and hybrid-ext models, respectively. This shows that from the point of view of expected neutrino events the hybrid-ext model is preferable. In an one-year exposure under the same rate of neutrino emission, which is a optimistic assumption as the rate was derived when the source was in an active state,  the expected events are $0.22$, $0.63$ and $1.15$ for the P-syn, hybrid and hybrid-ext models, respectively.

\section{Discussion and conclusions}\label{discon}

\pks\, is a bright blazar of a rather uncommon type that was found in spatial coincidence of IceCube, Baikal, Baksan and KM3NeT neutrinos detected in early December 2021 while it was undergoing its largest \gr, optical and soft X-ray flare observed since the launch of the Fermi satellite in 2008. Estimating a reliable value of the probability that this occurrence was due to a chance coincidence is however subject to uncertainties, being an "a posteriori" calculation.  
In order for \pks\, not to be related to \ic\, all the following occurrences must have been the result of concomitance due to random chance.
\begin{enumerate}
    \item a  bright BL Lac type blazar (2.2 to 5 Jy in the radio band) is located within the localization area of \ic\, slightly expanded from the nominal 90 percent error region to take into account systematic uncertainties (22 sq. deg). In the whole sky there are only seven BL Lac objects (one every 5,893 square degrees) with radio flux equal or larger than 2.2 Jy, as reported in BZCAT catalog 5th edition.
    \item the SED of \pks\, is of the IHBL type and is similar in shape and intensity to those of TXS\,0506+056, the source so far considered as the most likely neutrino candidate \citep{TXS0506Science2018}, as well as that of PKS\,1424+240 and GB6\,J1542+6129, the other two blazars found in correspondence to a significant excess in the IceCube 10-year neutrino sample \citep{IceCube10year,Abbasi2021}.
    \item we presented strong supporting evidence for \pks\,, like TXS 0506+056, PKS\,1424+240 and GB6\,J1542+6129, to be also a masquerading BL Lac. It also shares with these sources other features, including parsec scale properties, which are HSP-like but not
    FSRQ-like, redshift ($\sim 0.3 - 0.6$), and very high radio and $\gamma$-ray powers. \pks\, belongs in fact to the $\lesssim$ 1 per cent of the $\gamma$-ray selected blazars with such powers larger
    than those of TXS 0506+056 discussed by \cite{Padovani2022b}. 
    \item At the time of the arrival of the \ic\, neutrino, \pks\, was undergoing the largest \gr, X-ray and optical flare observed since 2008. Only four short and less intense flaring episodes are present in the over 13 year Fermi-LAT \gr\, light curve of this blazar (see Fig. \ref{lc}). 
    \item Three independent observatories detect additional neutrinos at different level of significance with position consistent with that of \pks\, within a short time of \ic\, and during the \gr\, flare. 

\end{enumerate}
Based on the above, \pks\, should be considered one of the best VHE neutrino source candidates detected so far. 
Thanks to the very good multiwavelength coverage from the optical to the HE band during the IceCube-211208A event, the available data set is very constraining for one-zone models. 
From the viewpoint of neutrino emission the most relevant parameters  are the maximum proton energy and the proton injection luminosity, $L_{\rm p, jet}$. In the P-syn model, the protons have to reach an energy of $\sim8.4\times10^{18}$ eV  
which is in the range of the observed energies of ultra-high-energy cosmic rays. 
However, this model characteristically produces a neutrino flux peaking  at much higher energies than that of IceCube-211208A~\citep[see also][for neutrino predictions in the P-syn model]{2020ApJ...893L..20L}. Instead, in the hybrid models, the hadronic contributions are limited by the X-ray data and assuming a moderate acceleration of protons up to $\gamma_{\rm p, max}=3.5\times10^5$ is enough to produce neutrinos at energies matching that of IceCube-211208A.  

A major difference between the applied models is the jet luminosity carried out by electrons, protons and the magnetic field. These luminosities 
are given in Table \ref{tabel_param}. In the P-syn model, the total jet luminosity is defined by the magnetic field ($L_{\rm B, jet}=3.81\times10^{47}\:{\rm erg\:s^{-1}}$) and proton content ($L_{\rm p, jet}=2.63\times10^{47}\:{\rm erg\:s^{-1}}$) and the emitting region is close to equipartition, ${\rm L_{\rm B, jet}/(L_{\rm p, jet}+L_{\rm e, jet}})\simeq1.82$. Still, the total power of a two-sided jet is $\sim 10^{48}$~erg s$^{-1}$, or up to 12.5 times higher than the Eddington luminosity of the source assuming a black hole mass $6.3\times10^8 M_{\odot}$ (note that $L_{\rm Edd} > 8 \times 10^{46}$ erg s$^{-1}$; Section \ref{sec:comparison})  for $z > 0.424$. 
This result is consistent with the findings of \citet{2020ApJ...893L..20L} who compared the minimum jet power in the P-syn scenario with the Eddington luminosity and the power of the Blandford–Znajek process for hundreds of blazars from the fourth Fermi AGN catalog \citep[4LAC,][]{2020ApJ...892..105A}.

The jet luminosity estimated in hybrid and hybrid-ext models differs as well. For the hybrid model, a substantially higher proton luminosity of
$L_{\rm p, jet}=1.36\times10^{50}\:{\rm erg\:s^{-1}}$ is obtained by
considering the X-ray observations and the emission from the secondary
pairs produced by the Bethe-Heitler process. Despite a much larger proton
luminosity for the hybrid model than for the P-syn model, the neutrino
luminosities are roughly comparable, even if the neutrino spectrum peaks
at lower energy for the hybrid model. This is because the maximum Lorentz
factor $\gamma_{\rm max}=3.5\times10^5$ is significantly lower than the
corresponding energy threshold of photopion production interactions with
the peak synchrotron photons, which reads
$\gamma_{\rm p}^{\rm (p\pi)}=1.3\times10^8$ assuming \nup$=5\times10^{15}$
Hz \citep[see Eq.~3 in][]{2015MNRAS.448.2412P}. As a result, protons
characterize by $\gamma_{\rm max}$ interact with higher-energy photons
which, however, have lower number density than the peak synchrotron photons.
This results in a reduced neutrino production efficiency, which is
here compensated by a higher proton luminosity. Nonetheless,
secondary electrons are efficiently produced from the Bethe-Heitler pair-production process, which has a lower energy threshold than the photopion production process. It is the synchrotron emission of these pairs that mostly constrains the proton luminosity by requiring that it does
not exceed the flux observed in the X-ray band. Instead, in the hybrid-ext model a modest $L_{\rm p, jet}=3.06\times10^{47}\:{\rm erg\:s^{-1}}$ luminosity was estimated. Similarly to the hybrid model, the proton contribution is limited by the radiation of the cascade in the X-ray band. Despite the lower proton density in the hybrid-ext model, the predicted neutrino flux is higher than for the hybrid model. This is due to an increase of the target photons provided by the BLR on which energetic protons interact.

All models applied here require a jet power that exceeds the Eddington luminosity estimated for \pks. Even though this is in agreement with the recent studies of neutrino emitting blazar candidates \citep[see e.g.,][]{2019MNRAS.483L..12C, 2019NatAs...3...88G,2018ApJ...864...84K}, such high powers are difficult to explain physically.
Note, however, that the jet power estimates were based on the modelling results of the brightest multi-wavelength flare detected from this source. As such, they should not be taken as representative of the long-term emission of \pks.

The baryon loading factor encodes information about the emission efficiency and it is defined as 
$\xi = L_{\rm p}/L_{\gamma}$, where $L_{\gamma}$ is the observed $\gamma$-ray luminosity integrated over the LAT energy range, and $L_{\rm p}=\delta^4 L'_{\rm p} \approx (4/3)\Gamma^2 L_{\rm p, jet}$. Using the values from Table~\ref{tabel_param} for the hybrid and hybrid-ext models, and $L_{\gamma}\simeq 3.9\times10^{47}$~erg s$^{-1}$ for the 21-day period of neutrino emission, we find $\xi \simeq 4\times10^5$ and $\xi \simeq 9\times10^2$, respectively. 
\cite{2020ApJ...899..113P} have shown in their Fig.~15 the baryon loading factors obtained by SED modelling of BL Lacs, including TXS~0506+056 with a similar hybrid emission model. Our $\xi$ value for the hybrid model is closer to radiation models of BL Lacs, where only the jet synchrotron photons are targets for photohadronic interactions, and is about two orders of magnitude larger than the one found by \cite{2018ApJ...864...84K} for TXS~0506+056 during its 2017 multi-frequency flare.
This difference is because in the hybrid model adopted by \cite{2018ApJ...864...84K}, neutrino production is mostly achieved via photo-pion interactions on external radiation fields, which is similar to the hybrid-ext model examined here. Indeed, the $\xi$ value for  our hybrid-ext model  is very close to the value estimated for TXS~0506+056 during its 2017 multi-frequency flare.
The ratio of the all-flavour neutrino luminosity to the $\gamma$-ray luminosity for \pks\, is $L_{\nu+\bar{\nu}}/L_{\gamma}\simeq3.3\times10^{-2}$ for hybrid model and $\simeq6.9\times10^{-3}$ for hybrid-ext model, which are also comparable to the values found for TXS 0506+056 during the 2017 flare \cite[see the red markers in Fig.~15 in][]{2020ApJ...899..113P}. 

We next discuss some alternative models for predicting the neutrino output of \pks\ during its most recent multiwavelength flare. We first consider a scenario where X-ray flares are powered by synchrotron radiation of relativistic protons, and VHE neutrinos are produced through photomeson interactions between protons with their own synchrotron X-ray photons~\citep{2021ApJ...906..131M}. Following the methodology described in \citet{2022MNRAS.510.4063S}, we identify flaring states using the Bayesian block representation of the 1 keV X-ray light curve shown in Fig.~\ref{lcmulti}. Then, using the 0.5-10 keV fluence of each flaring state as a proxy for the all-flavour neutrino fluence, and the IceCube point-source effective area  for the declination of \pks\, \citep{IceCube10year}, we calculate the number of muon and antimuon neutrinos above 100 TeV expected for IceCube. The predicted neutrino flux is similar to that found in the hybrid model, as it is limited by the X-ray flux in both scenarios. The peak neutrino energy in the hadronic X-ray flaring scenario also falls in the 0.1-1 PeV range -- see Eq.~(3) in  \citet{2022MNRAS.510.4063S}. 
We predict $N_{\nu_{\mu}+\bar{\nu}_{\mu}}= 0.008 \pm 0.002$ for a total flaring duration of 8.61 days. Assuming a similar flaring duration as for the \gr \ (i.e. 21.3 days), we expect $\sim 0.0120\pm 0.005$ muon and anti-muon neutrinos. The prediction of this model is closer to the one from the hybrid model, since the neutrino fluence in both scenarios is similar to the X-ray fluence.

Given that the source was found to be flaring simultaneously in optical/UV and X-ray wavelengths, one might wonder if the hadronic flare scenario could be applied to the lower energy data. Let us assume that proton synchrotron radiation was responsible for producing photons with energy as low as $1$~eV. Then, extremely high proton energies would be required to meet the threshold condition for photopion production on the same proton synchrotron photons, namely $\gamma_{\rm p}\gtrsim 1.5\times10^9 (\delta/10)$~\citep[see Eqs.~1 and 3 in][]{2021ApJ...906..131M}. At the same time, the magnetic field strength required to produce 1~eV synchrotron photons would be extremely low, namely $B\lesssim 7 \, (\delta/10)^{-3}$~nG~\citep[see Eq.~1 in][]{2021ApJ...906..131M}. Moreover, this model would predict neutrinos with observed energies $\gtrsim 0.75~(\delta/10)^2$~EeV. It is therefore unlikely that a hadronic scenario for the UV/X-ray flare would explain the tentative association with a $\sim 100$~TeV neutrino like \ic.

The models applied in this paper assume significant neutrino emission above $\sim100$ TeV while predicting a very low flux of sub-TeV neutrinos (see Figs.~\ref{sed_model} and \ref{sed_hybrid_ext}). Instead, alternative hadronic models that assume accelerated protons interact with a dense target crossing the jet (cloud, star envelope, etc.) also predict low-energy neutrino emission, which might explain the events observed by Baikal and Baksan instruments (if they are associated with \pks). In these models the protons should be accelerated only up to moderate energies and their contribution is mostly released in sub-TeV band. Moreover, in \citet{2018ApJ...866..109S} and \citet{2019PhRvD..99f3008L} such inelastic $pp$ interaction scenario was applied to model VHE neutrinos from the direction of TXS 0506+056. The modeling of the multi-wavelength and GeV-TeV neutrino emission from \pks\ will be addressed in a future study.

In this work we have compared the predictions of a range of physical scenarios for neutrino production to  the rich multi-frequency data set 
available for \pks. All models predict a number of IceCube neutrinos that is 
less than 1 during the Dec 2021 flare, with the largest expectation reaching 0.067 for the case of the hybrid-ext model. This result may be regarded as  unsatisfactory, although the Poisson probability of observing one neutrino when the expectation is 0.067 is $\sim6.7$\%, which is a non-negligible probability, statistically consistent with the observations. 
We note that if the models predicted a number of IceCube neutrinos of the order of 1.0, then,  we would have a larger problem of consistency with other observational data. Since the SED of PKS0735+178 during the Dec 2021 flare is very similar to that of TXS\,0506+056 (see Fig. \ref{sedscomparison}), PKS 1424+240 and GB6 J1542+6129  \citep{GiommiPadovani2021} a prediction of the order of one IceCube neutrino during a flare or a  high state should apply to these sources as well and should be extended to all similar blazars when flaring. It is not straightforward to estimate how many flaring events happened over the past few years in 
IHBLs. For example, in the Swift-XRT blazar database of \cite{Giommi2019} more than 100 IHBL blazars at least once were detected in the flaring state with a soft X-ray flux similar or larger  than that of \pks\ in Dec 2021.
Although a database of \gr\ flares in blazars is not available it is reasonable to assume that a similar number of  IHBL sources flared to \gr\, fluxes similar to that of \pks. Therefore the number of flaring blazars in the X-ray and \gr\ bands should be relatively large and the estimation of $\sim 1 $ IceCube neutrino for each flaring event would then predict the detection of at least several dozens of IceCube neutrino track events in spatial coincidence with bright IHBL blazars. Clearly this is not the case, in agreement with the low prediction of the models considered here.

A \gr\, flux similar to that observed during the December 2021 flare was recorded by the EGRET detector \citep{EGRET} in the 1990’s. In the optical band also, the source was very bright in 1976 \citep{2007A&A...467..465C}. 
\pks\, can therefore be expected to brighten again sometime in the near-mid future and, if the \gr\, or optical flux is a good proxy for neutrino emission, more neutrinos might be detected.  
As currently operating neutrino observatories are bound to improve their sensitivity over the next few years and other facilities, like P-One \citep{PONE}, 
KM3NeT\footnote{\url{https://www.km3net.org/}} and 
IceCube-Gen2\footnote{\url{https://icecube.wisc.edu/science/beyond/}}, are expected to come on-line, if the multi-messenger flare observed in December 2021 is not just a large statistical fluctuation, many more neutrinos from \pks\, and similar sources should be detected during flares, making blazars major targets for the next phase of neutrino and multi-messenger astrophysics.

\section*{Acknowledgements}
The authors would like to thank Stamatios I. Stathopoulos for the neutrino estimations in the hadronic flare model and Simona Paiano for her help with the upper limits 
on the \ion{O}{II} flux and for a useful discussion.
We acknowledge the use of data, analysis tools and services from the Open Universe platform, the ASI Space Science Data Center (SSDC), the Astrophysics Science Archive Research Center (HEASARC), the Astrophysics Data System (ADS), the National Extra-galactic Database (NED), and the ASAS-SN sky server.

NS acknowledges the support by the Science Committee of RA, in the frames of the research project No 20TTCG-1C015.

MP acknowledges support from the MERAC
Foundation through the project THRILL. 

SG acknowledges the support by the Science Committee of RA, in the frames of the research project No 21T-1C260.

DB acknowledges support from the European Research Council via the ERC consolidating grant $\sharp$773062 (acronym O.M.J.).

\section*{Data Availability}
The data underlying this article will be shared on reasonable request to the corresponding author.



\bibliographystyle{mnras}
\bibliography{PKS0735} 

\begin{thebibliography}{}
\makeatletter
\relax
\def\mn@urlcharsother{\let\do\@makeother \do\$\do\&\do\#\do\^\do\_\do\%\do\~}
\def\mn@doi{\begingroup\mn@urlcharsother \@ifnextchar [ {\mn@doi@}
  {\mn@doi@[]}}
\def\mn@doi@[#1]#2{\def\@tempa{#1}\ifx\@tempa\@empty \href
  {http://dx.doi.org/#2} {doi:#2}\else \href {http://dx.doi.org/#2} {#1}\fi
  \endgroup}
\def\mn@eprint#1#2{\mn@eprint@#1:#2::\@nil}
\def\mn@eprint@arXiv#1{\href {http://arxiv.org/abs/#1} {{\tt arXiv:#1}}}
\def\mn@eprint@dblp#1{\href {http://dblp.uni-trier.de/rec/bibtex/#1.xml}
  {dblp:#1}}
\def\mn@eprint@#1:#2:#3:#4\@nil{\def\@tempa {#1}\def\@tempb {#2}\def\@tempc
  {#3}\ifx \@tempc \@empty \let \@tempc \@tempb \let \@tempb \@tempa \fi \ifx
  \@tempb \@empty \def\@tempb {arXiv}\fi \@ifundefined
  {mn@eprint@\@tempb}{\@tempb:\@tempc}{\expandafter \expandafter \csname
  mn@eprint@\@tempb\endcsname \expandafter{\@tempc}}}

\bibitem[\protect\citeauthoryear{{Aartsen}, {Abbasi}, {Abdou}, {Ackermann},
  {Adams}, {Aguilar}  \& et al.}{{Aartsen} et~al.}{2013}]{Aartsen2013}
{Aartsen} M.~G.,  {Abbasi} R.,  {Abdou} Y.,  {Ackermann} M.,  {Adams} J.,
  {Aguilar} J.~A.,   et al. 2013, \mn@doi [\prl]
  {10.1103/PhysRevLett.111.021103}, \href
  {https://ui.adsabs.harvard.edu/abs/2013PhRvL.111b1103A} {111, 021103}

\bibitem[\protect\citeauthoryear{{Aartsen}, {Ackermann}, {Adams}, {Aguilar},
  {Ahlers}, {Ahrens}, {Alispach}  \& et al.}{{Aartsen}
  et~al.}{2020}]{IceCube10year}
{Aartsen} M.~G.,  {Ackermann} M.,  {Adams} J.,  {Aguilar} J.~A.,  {Ahlers} M.,
  {Ahrens} M.,  {Alispach} C.,   et al. 2020, \mn@doi [\prl]
  {10.1103/PhysRevLett.124.051103}, \href
  {https://ui.adsabs.harvard.edu/abs/2020PhRvL.124e1103A} {124, 051103}

\bibitem[\protect\citeauthoryear{{Abdo}, {Ackermann}, {Agudo}, {Ajello},
  {Aller}, {Aller}, {Angelakis}  \& et al.}{{Abdo} et~al.}{2010}]{Abdo2010}
{Abdo} A.~A.,  {Ackermann} M.,  {Agudo} I.,  {Ajello} M.,  {Aller} H.~D.,
  {Aller} M.~F.,  {Angelakis} E.,   et al. 2010, \mn@doi [\apj]
  {10.1088/0004-637X/716/1/30}, \href
  {https://ui.adsabs.harvard.edu/abs/2010ApJ...716...30A} {716, 30}

\bibitem[\protect\citeauthoryear{{Agostini}, {B{\"o}hmer}, {Bosma}, {Clark},
  {Danninger}, {Fruck}, {Gernh{\"a}user}  \& et al.}{{Agostini}
  et~al.}{2020}]{PONE}
{Agostini} M.,  {B{\"o}hmer} M.,  {Bosma} J.,  {Clark} K.,  {Danninger} M.,
  {Fruck} C.,  {Gernh{\"a}user} R.,   et al. 2020, \mn@doi [Nature Astronomy]
  {10.1038/s41550-020-1182-4}, \href
  {https://ui.adsabs.harvard.edu/abs/2020NatAs...4..913A} {4, 913}

\bibitem[\protect\citeauthoryear{{Ajello} et~al.,}{{Ajello}
  et~al.}{2020}]{2020ApJ...892..105A}
{Ajello} M.,  et~al., 2020, \mn@doi [\apj] {10.3847/1538-4357/ab791e}, \href
  {https://ui.adsabs.harvard.edu/abs/2020ApJ...892..105A} {892, 105}

\bibitem[\protect\citeauthoryear{{Britzen} et~al.,}{{Britzen}
  et~al.}{2010}]{Britzen2010}
{Britzen} S.,  et~al., 2010, \mn@doi [\aap] {10.1051/0004-6361/200913685},
  \href {https://ui.adsabs.harvard.edu/abs/2010A&A...515A.105B} {515, A105}

\bibitem[\protect\citeauthoryear{{Carrasco}, {Recillas}, {Escobedo}, {Porras},
  {Chavushyan}  \& {Mayya}}{{Carrasco} et~al.}{2021}]{ATel15148}
{Carrasco} L.,  {Recillas} E.,  {Escobedo} G.,  {Porras} A.,  {Chavushyan} V.,
   {Mayya} Y.~D.,  2021, The Astronomer's Telegram, \href
  {https://ui.adsabs.harvard.edu/abs/2021ATel15148....1C} {15148, 1}

\bibitem[\protect\citeauthoryear{{Carswell}, {Strittmatter}, {Williams},
  {Kinman}  \& {Serkowski}}{{Carswell} et~al.}{1974}]{Carswell1974}
{Carswell} R.~F.,  {Strittmatter} P.~A.,  {Williams} R.~E.,  {Kinman} T.~D.,
  {Serkowski} K.,  1974, \mn@doi [\apjl] {10.1086/181516}, \href
  {https://ui.adsabs.harvard.edu/abs/1974ApJ...190L.101C} {190, L101}

\bibitem[\protect\citeauthoryear{{Cerruti}, {Zech}, {Boisson}  \&
  {Inoue}}{{Cerruti} et~al.}{2015}]{2015MNRAS.448..910C}
{Cerruti} M.,  {Zech} A.,  {Boisson} C.,   {Inoue} S.,  2015, \mn@doi [\mnras]
  {10.1093/mnras/stu2691}, \href
  {https://ui.adsabs.harvard.edu/abs/2015MNRAS.448..910C} {448, 910}

\bibitem[\protect\citeauthoryear{{Cerruti}, {Zech}, {Boisson}, {Emery}, {Inoue}
   \& {Lenain}}{{Cerruti} et~al.}{2019}]{2019MNRAS.483L..12C}
{Cerruti} M.,  {Zech} A.,  {Boisson} C.,  {Emery} G.,  {Inoue} S.,   {Lenain}
  J.~P.,  2019, \mn@doi [\mnras] {10.1093/mnrasl/sly210}, \href
  {https://ui.adsabs.harvard.edu/abs/2019MNRAS.483L..12C} {483, L12}

\bibitem[\protect\citeauthoryear{{Chang}, {Brandt}  \& {Giommi}}{{Chang}
  et~al.}{2020}]{voublazars}
{Chang} Y.~L.,  {Brandt} C.~H.,   {Giommi} P.,  2020, \mn@doi [Astronomy and
  Computing] {10.1016/j.ascom.2019.100350}, \href
  {https://ui.adsabs.harvard.edu/abs/2020A&C....3000350C} {30, 100350}

\bibitem[\protect\citeauthoryear{{Ciprini} et~al.,}{{Ciprini}
  et~al.}{2007}]{2007A&A...467..465C}
{Ciprini} S.,  et~al., 2007, \mn@doi [\aap] {10.1051/0004-6361:20052646}, \href
  {https://ui.adsabs.harvard.edu/abs/2007A&A...467..465C} {467, 465}

\bibitem[\protect\citeauthoryear{Condon, Cotton, Greisen, Yin, Perley, Taylor
  \& Broderick}{Condon et~al.}{1998}]{Condon1998}
Condon J.,  Cotton W.,  Greisen E.,  Yin Q.,  Perley R.,  Taylor G.,
  Broderick J.,  1998, \mn@doi [AJ] {10.1086/300337}, 115, 1693

\bibitem[\protect\citeauthoryear{{Dimitrakoudis}, {Petropoulou}  \&
  {Mastichiadis}}{{Dimitrakoudis} et~al.}{2014}]{2014APh....54...61D}
{Dimitrakoudis} S.,  {Petropoulou} M.,   {Mastichiadis} A.,  2014, \mn@doi
  [Astroparticle Physics] {10.1016/j.astropartphys.2013.10.005}, \href
  {https://ui.adsabs.harvard.edu/abs/2014APh....54...61D} {54, 61}

\bibitem[\protect\citeauthoryear{{Dom{\'\i}nguez} et~al.,}{{Dom{\'\i}nguez}
  et~al.}{2011}]{2011MNRAS.410.2556D}
{Dom{\'\i}nguez} A.,  et~al., 2011, \mn@doi [\mnras]
  {10.1111/j.1365-2966.2010.17631.x}, \href
  {https://ui.adsabs.harvard.edu/abs/2011MNRAS.410.2556D} {410, 2556}

\bibitem[\protect\citeauthoryear{{Dzhilkibaev}, {Suvorova}  \& {Baikal-GVD
  Collaboration}}{{Dzhilkibaev} et~al.}{2021}]{Baikal}
{Dzhilkibaev} Z.~A.,  {Suvorova} O.,   {Baikal-GVD Collaboration} 2021, The
  Astronomer's Telegram, \href
  {https://ui.adsabs.harvard.edu/abs/2021ATel15112....1D} {15112, 1}

\bibitem[\protect\citeauthoryear{{Falomo} \& {Ulrich}}{{Falomo} \&
  {Ulrich}}{2000}]{2000A&A...357...91F}
{Falomo} R.,  {Ulrich} M.~H.,  2000, \aap, \href
  {https://ui.adsabs.harvard.edu/abs/2000A&A...357...91F} {357, 91}

\bibitem[\protect\citeauthoryear{{Falomo}, {Pian}  \& {Treves}}{{Falomo}
  et~al.}{2014}]{Falomo2014}
{Falomo} R.,  {Pian} E.,   {Treves} A.,  2014, \mn@doi [\aapr]
  {10.1007/s00159-014-0073-z}, \href
  {https://ui.adsabs.harvard.edu/abs/2014A&ARv..22...73F} {22, 73}

\bibitem[\protect\citeauthoryear{{Falomo}, {Treves}  \& {Paiano}}{{Falomo}
  et~al.}{2021}]{Falomo2021}
{Falomo} R.,  {Treves} A.,   {Paiano} S.,  2021, The Astronomer's Telegram,
  \href {https://ui.adsabs.harvard.edu/abs/2021ATel15132....1F} {15132, 1}

\bibitem[\protect\citeauthoryear{{Fanaroff} \& {Riley}}{{Fanaroff} \&
  {Riley}}{1974}]{Fanaroff1974}
{Fanaroff} B.~L.,  {Riley} J.~M.,  1974, \mn@doi [\mnras]
  {10.1093/mnras/167.1.31P}, \href
  {https://ui.adsabs.harvard.edu/abs/1974MNRAS.167P..31F} {167, 31P}

\bibitem[\protect\citeauthoryear{{Fanidakis}, {Baugh}, {Benson}, {Bower},
  {Cole}, {Done}  \& {Frenk}}{{Fanidakis} et~al.}{2011}]{Fanidakis2011}
{Fanidakis} N.,  {Baugh} C.~M.,  {Benson} A.~J.,  {Bower} R.~G.,  {Cole} S.,
  {Done} C.,   {Frenk} C.~S.,  2011, \mn@doi [\mnras]
  {10.1111/j.1365-2966.2010.17427.x}, \href
  {https://ui.adsabs.harvard.edu/abs/2011MNRAS.410...53F} {410, 53}

\bibitem[\protect\citeauthoryear{{Feng}, {Jin}, {Mori}, {Mukherjee},
  {Santander}  \& {Woo}}{{Feng} et~al.}{2021}]{ATel15113}
{Feng} Q.,  {Jin} W.,  {Mori} K.,  {Mukherjee} R.,  {Santander} M.,   {Woo} J.,
   2021, The Astronomer's Telegram, \href
  {https://ui.adsabs.harvard.edu/abs/2021ATel15113....1F} {15113, 1}

\bibitem[\protect\citeauthoryear{{Filippini} et~al.,}{{Filippini}
  et~al.}{2022}]{ATel15290}
{Filippini} F.,  et~al., 2022, The Astronomer's Telegram, \href
  {https://ui.adsabs.harvard.edu/abs/2022ATel15290....1F} {15290, 1}

\bibitem[\protect\citeauthoryear{{Gao}, {Fedynitch}, {Winter}  \& {Pohl}}{{Gao}
  et~al.}{2019}]{2019NatAs...3...88G}
{Gao} S.,  {Fedynitch} A.,  {Winter} W.,   {Pohl} M.,  2019, \mn@doi [Nature
  Astronomy] {10.1038/s41550-018-0610-1}, \href
  {https://ui.adsabs.harvard.edu/abs/2019NatAs...3...88G} {3, 88}

\bibitem[\protect\citeauthoryear{{Gasparyan}, {B{\'e}gu{\'e}}  \&
  {Sahakyan}}{{Gasparyan} et~al.}{2022}]{2022MNRAS.509.2102G}
{Gasparyan} S.,  {B{\'e}gu{\'e}} D.,   {Sahakyan} N.,  2022, \mn@doi [\mnras]
  {10.1093/mnras/stab2688}, \href
  {https://ui.adsabs.harvard.edu/abs/2022MNRAS.509.2102G} {509, 2102}

\bibitem[\protect\citeauthoryear{{Gehrels} et~al.,}{{Gehrels}
  et~al.}{2004}]{SwiftPaper}
{Gehrels} N.,  et~al., 2004, \mn@doi [\apj] {10.1086/422091}, \href
  {https://ui.adsabs.harvard.edu/abs/2004ApJ...611.1005G} {611, 1005}

\bibitem[\protect\citeauthoryear{{Ghisellini} \& {Madau}}{{Ghisellini} \&
  {Madau}}{1996}]{1996MNRAS.280...67G}
{Ghisellini} G.,  {Madau} P.,  1996, \mn@doi [\mnras] {10.1093/mnras/280.1.67},
  \href {https://ui.adsabs.harvard.edu/abs/1996MNRAS.280...67G} {280, 67}

\bibitem[\protect\citeauthoryear{{Ghisellini} \& {Tavecchio}}{{Ghisellini} \&
  {Tavecchio}}{2008}]{2008MNRAS.387.1669G}
{Ghisellini} G.,  {Tavecchio} F.,  2008, \mn@doi [\mnras]
  {10.1111/j.1365-2966.2008.13360.x}, \href
  {https://ui.adsabs.harvard.edu/abs/2008MNRAS.387.1669G} {387, 1669}

\bibitem[\protect\citeauthoryear{{Ghisellini}, {Tavecchio}, {Foschini}  \&
  {Ghirlanda}}{{Ghisellini} et~al.}{2011}]{Ghisellini2011}
{Ghisellini} G.,  {Tavecchio} F.,  {Foschini} L.,   {Ghirlanda} G.,  2011,
  \mn@doi [\mnras] {10.1111/j.1365-2966.2011.18578.x}, \href
  {https://ui.adsabs.harvard.edu/abs/2011MNRAS.414.2674G} {414, 2674}

\bibitem[\protect\citeauthoryear{{Giommi} \& {Padovani}}{{Giommi} \&
  {Padovani}}{2021}]{GiommiPadovani2021}
{Giommi} P.,  {Padovani} P.,  2021, \mn@doi [Universe]
  {10.3390/universe7120492}, \href
  {https://ui.adsabs.harvard.edu/abs/2021Univ....7..492G} {7, 492}

\bibitem[\protect\citeauthoryear{{Giommi}, {Barr}, {Garilli}, {Maccagni}  \&
  {Pollock}}{{Giommi} et~al.}{1990}]{Giommi1990}
{Giommi} P.,  {Barr} P.,  {Garilli} B.,  {Maccagni} D.,   {Pollock} A.~M.~T.,
  1990, \mn@doi [\apj] {10.1086/168851}, \href
  {https://ui.adsabs.harvard.edu/abs/1990ApJ...356..432G} {356, 432}

\bibitem[\protect\citeauthoryear{{Giommi} et~al.,}{{Giommi}
  et~al.}{2019}]{Giommi2019}
{Giommi} P.,  et~al., 2019, \mn@doi [\aap] {10.1051/0004-6361/201935646}, \href
  {https://ui.adsabs.harvard.edu/abs/2019A&A...631A.116G} {631, A116}

\bibitem[\protect\citeauthoryear{{Giommi} et~al.,}{{Giommi}
  et~al.}{2020a}]{GiommiOU}
{Giommi} P.,  et~al., 2020a, in S. F.,  ed., Space Capacity Building in the XXI
  Century. Studies in Space Policy. Springer, pp 377--386 (\mn@eprint {arXiv}
  {1805.08505}), \mn@doi{10.1007/978-3-030-21938-3}

\bibitem[\protect\citeauthoryear{{Giommi}, {Glauch}, {Padovani}, {Resconi},
  {Turcati}  \& {Chang}}{{Giommi} et~al.}{2020b}]{Giommidissecting}
{Giommi} P.,  {Glauch} T.,  {Padovani} P.,  {Resconi} E.,  {Turcati} A.,
  {Chang} Y.~L.,  2020b, \mn@doi [\mnras] {10.1093/mnras/staa2082}, \href
  {https://ui.adsabs.harvard.edu/abs/2020MNRAS.497..865G} {497, 865}

\bibitem[\protect\citeauthoryear{{Giommi} et~al.,}{{Giommi}
  et~al.}{2021}]{Giommi2021}
{Giommi} P.,  et~al., 2021, \mn@doi [\mnras] {10.1093/mnras/stab2425}, \href
  {https://ui.adsabs.harvard.edu/abs/2021MNRAS.507.5690G} {507, 5690}

\bibitem[\protect\citeauthoryear{{Haemmerich}, {Zainab}, {Gokus}, {Weber},
  {Kreykenbohm}  \& {Wilms}}{{Haemmerich} et~al.}{2021}]{ATel15108}
{Haemmerich} S.,  {Zainab} A.,  {Gokus} A.,  {Weber} P.,  {Kreykenbohm} I.,
  {Wilms} J.,  2021, The Astronomer's Telegram, \href
  {https://ui.adsabs.harvard.edu/abs/2021ATel15108....1H} {15108, 1}

\bibitem[\protect\citeauthoryear{{Halzen} \& {Zas}}{{Halzen} \&
  {Zas}}{1997}]{HalzeZas1997}
{Halzen} F.,  {Zas} E.,  1997, \mn@doi [\apj] {10.1086/304741}, \href
  {https://ui.adsabs.harvard.edu/abs/1997ApJ...488..669H} {488, 669}

\bibitem[\protect\citeauthoryear{{Hartman} et~al.,}{{Hartman}
  et~al.}{1999}]{EGRET}
{Hartman} R.~C.,  et~al., 1999, \mn@doi [\apjs] {10.1086/313231}, \href
  {https://ui.adsabs.harvard.edu/abs/1999ApJS..123...79H} {123, 79}

\bibitem[\protect\citeauthoryear{{Henri} \& {Saug{\'e}}}{{Henri} \&
  {Saug{\'e}}}{2006}]{2006ApJ...640..185H}
{Henri} G.,  {Saug{\'e}} L.,  2006, \mn@doi [\apj] {10.1086/500039}, \href
  {https://ui.adsabs.harvard.edu/abs/2006ApJ...640..185H} {640, 185}

\bibitem[\protect\citeauthoryear{{Homan} et~al.,}{{Homan}
  et~al.}{2021}]{Homan2021}
{Homan} D.~C.,  et~al., 2021, \mn@doi [\apj] {10.3847/1538-4357/ac27af}, \href
  {https://ui.adsabs.harvard.edu/abs/2021ApJ...923...67H} {923, 67}

\bibitem[\protect\citeauthoryear{{Hovatta} et~al.,}{{Hovatta}
  et~al.}{2021}]{Hovatta2021}
{Hovatta} T.,  et~al., 2021, \mn@doi [\aap] {10.1051/0004-6361/202039481},
  \href {https://ui.adsabs.harvard.edu/abs/2021A&A...650A..83H} {650, A83}

\bibitem[\protect\citeauthoryear{{IceCube Collaboration}}{{IceCube
  Collaboration}}{2018}]{TXS0506Science2018}
{IceCube Collaboration} 2018, \mn@doi [Science] {10.1126/science.aat1378},
  \href {https://ui.adsabs.harvard.edu/abs/2018Sci...361.1378I} {361, eaat1378}

\bibitem[\protect\citeauthoryear{{IceCube Collaboration}}{{IceCube
  Collaboration}}{2021a}]{Abbasi2021}
{IceCube Collaboration} 2021a, \mn@doi [\apjl] {10.3847/2041-8213/ac2c7b},
  \href {https://ui.adsabs.harvard.edu/abs/2021ApJ...920L..45A} {920, L45}

\bibitem[\protect\citeauthoryear{{IceCube Collaboration}}{{IceCube
  Collaboration}}{2021b}]{NGC1068GCN}
{IceCube Collaboration} 2021b, GRB Coordinates Network, \href
  {https://ui.adsabs.harvard.edu/abs/2021GCN.31085....1I} {31085, 1}

\bibitem[\protect\citeauthoryear{{IceCube Collaboration}}{{IceCube
  Collaboration}}{2021c}]{GCN3119}
{IceCube Collaboration} 2021c, GRB Coordinates Network, \href
  {https://ui.adsabs.harvard.edu/abs/2021GCN.31191....1I} {31191, 1}

\bibitem[\protect\citeauthoryear{{Impey} \& {Tapia}}{{Impey} \&
  {Tapia}}{1990}]{ImpeyTapia1990}
{Impey} C.~D.,  {Tapia} S.,  1990, \mn@doi [\apj] {10.1086/168672}, \href
  {https://ui.adsabs.harvard.edu/abs/1990ApJ...354..124I} {354, 124}

\bibitem[\protect\citeauthoryear{{Jannuzi}, {Smith}  \& {Elston}}{{Jannuzi}
  et~al.}{1994}]{Jannuzi1994}
{Jannuzi} B.~T.,  {Smith} P.~S.,   {Elston} R.,  1994, \mn@doi [\apj]
  {10.1086/174226}, \href
  {https://ui.adsabs.harvard.edu/abs/1994ApJ...428..130J} {428, 130}

\bibitem[\protect\citeauthoryear{{Jorstad} et~al.,}{{Jorstad}
  et~al.}{2017}]{Jorstad2017}
{Jorstad} S.~G.,  et~al., 2017, \mn@doi [\apj] {10.3847/1538-4357/aa8407},
  \href {https://ui.adsabs.harvard.edu/abs/2017ApJ...846...98J} {846, 98}

\bibitem[\protect\citeauthoryear{{Kadler} et~al.,}{{Kadler}
  et~al.}{2021}]{Atel15105}
{Kadler} M.,  et~al., 2021, The Astronomer's Telegram, \href
  {https://ui.adsabs.harvard.edu/abs/2021ATel15105....1K} {15105, 1}

\bibitem[\protect\citeauthoryear{{Keivani} et~al.,}{{Keivani}
  et~al.}{2018}]{2018ApJ...864...84K}
{Keivani} A.,  et~al., 2018, \mn@doi [\apj] {10.3847/1538-4357/aad59a}, \href
  {https://ui.adsabs.harvard.edu/abs/2018ApJ...864...84K} {864, 84}

\bibitem[\protect\citeauthoryear{{Kochanek} et~al.,}{{Kochanek}
  et~al.}{2017}]{ASAS-SN}
{Kochanek} C.~S.,  et~al., 2017, \mn@doi [\pasp] {10.1088/1538-3873/aa80d9},
  \href {https://ui.adsabs.harvard.edu/abs/2017PASP..129j4502K} {129, 104502}

\bibitem[\protect\citeauthoryear{{Labita}, {Treves}, {Falomo}  \&
  {Uslenghi}}{{Labita} et~al.}{2006}]{Labita2006}
{Labita} M.,  {Treves} A.,  {Falomo} R.,   {Uslenghi} M.,  2006, \mn@doi
  [\mnras] {10.1111/j.1365-2966.2006.10878.x}, \href
  {https://ui.adsabs.harvard.edu/abs/2006MNRAS.373..551L} {373, 551}

\bibitem[\protect\citeauthoryear{{Lagunas Gualda}, {Ashida}, {Sharma}  \&
  {Thomas}}{{Lagunas Gualda} et~al.}{2021}]{IceCubeSystematics}
{Lagunas Gualda} C.,  {Ashida} Y.,  {Sharma} A.,   {Thomas} H.,  2021, arXiv
  e-prints, \href {https://ui.adsabs.harvard.edu/abs/2021arXiv210708670L} {p.
  arXiv:2107.08670}

\bibitem[\protect\citeauthoryear{{Lindfors} et~al.,}{{Lindfors}
  et~al.}{2021}]{ATel15136}
{Lindfors} E.,  et~al., 2021, The Astronomer's Telegram, \href
  {https://ui.adsabs.harvard.edu/abs/2021ATel15136....1L} {15136, 1}

\bibitem[\protect\citeauthoryear{{Liodakis} \& {Petropoulou}}{{Liodakis} \&
  {Petropoulou}}{2020}]{2020ApJ...893L..20L}
{Liodakis} I.,  {Petropoulou} M.,  2020, \mn@doi [\apjl]
  {10.3847/2041-8213/ab830a}, \href
  {https://ui.adsabs.harvard.edu/abs/2020ApJ...893L..20L} {893, L20}

\bibitem[\protect\citeauthoryear{{Lister} et~al.,}{{Lister}
  et~al.}{2019}]{Lister2019}
{Lister} M.~L.,  et~al., 2019, \mn@doi [\apj] {10.3847/1538-4357/ab08ee}, \href
  {https://ui.adsabs.harvard.edu/abs/2019ApJ...874...43L} {874, 43}

\bibitem[\protect\citeauthoryear{{Liu}, {Wang}, {Xue}, {Taylor}, {Wang}, {Li}
  \& {Yan}}{{Liu} et~al.}{2019}]{2019PhRvD..99f3008L}
{Liu} R.-Y.,  {Wang} K.,  {Xue} R.,  {Taylor} A.~M.,  {Wang} X.-Y.,  {Li} Z.,
  {Yan} H.,  2019, \mn@doi [\prd] {10.1103/PhysRevD.99.063008}, \href
  {https://ui.adsabs.harvard.edu/abs/2019PhRvD..99f3008L} {99, 063008}

\bibitem[\protect\citeauthoryear{{Lott}, {Escande}, {Larsson}  \&
  {Ballet}}{{Lott} et~al.}{2012}]{2012A&A...544A...6L}
{Lott} B.,  {Escande} L.,  {Larsson} S.,   {Ballet} J.,  2012, \mn@doi [\aap]
  {10.1051/0004-6361/201218873}, \href
  {https://ui.adsabs.harvard.edu/abs/2012A&A...544A...6L} {544, A6}

\bibitem[\protect\citeauthoryear{{Lott}, {Gasparrini}  \& {Ciprini}}{{Lott}
  et~al.}{2020}]{4LAC-DR2}
{Lott} B.,  {Gasparrini} D.,   {Ciprini} S.,  2020, arXiv e-prints, \href
  {https://ui.adsabs.harvard.edu/abs/2020arXiv201008406L} {p. arXiv:2010.08406}

\bibitem[\protect\citeauthoryear{{Madejski} \& {Schwartz}}{{Madejski} \&
  {Schwartz}}{1988}]{Madejski1988}
{Madejski} G.~M.,  {Schwartz} D.~A.,  1988, \mn@doi [\apj] {10.1086/166511},
  \href {https://ui.adsabs.harvard.edu/abs/1988ApJ...330..776M} {330, 776}

\bibitem[\protect\citeauthoryear{{Mannheim}}{{Mannheim}}{1993}]{Mannheim1993}
{Mannheim} K.,  1993, \mn@doi [\prd] {10.1103/PhysRevD.48.2408}, \href
  {https://ui.adsabs.harvard.edu/abs/1993PhRvD..48.2408M} {48, 2408}

\bibitem[\protect\citeauthoryear{Massaro, Maselli, Leto, Marchegiani, Perri,
  Giommi  \& Piranomonte}{Massaro et~al.}{2015}]{Massaro2015}
Massaro E.,  Maselli A.,  Leto C.,  Marchegiani P.,  Perri M.,  Giommi P.,
  Piranomonte S.,  2015, \mn@doi [Ap{\&}SS] {10.1007/s10509-015-2254-2}, 357,
  75

\bibitem[\protect\citeauthoryear{{Mastichiadis} \& {Moraitis}}{{Mastichiadis}
  \& {Moraitis}}{2008}]{2008A&A...491L..37M}
{Mastichiadis} A.,  {Moraitis} K.,  2008, \mn@doi [\aap]
  {10.1051/0004-6361:200810505}, \href
  {https://ui.adsabs.harvard.edu/abs/2008A&A...491L..37M} {491, L37}

\bibitem[\protect\citeauthoryear{{Mastichiadis} \&
  {Petropoulou}}{{Mastichiadis} \& {Petropoulou}}{2021}]{2021ApJ...906..131M}
{Mastichiadis} A.,  {Petropoulou} M.,  2021, \mn@doi [\apj]
  {10.3847/1538-4357/abc952}, \href
  {https://ui.adsabs.harvard.edu/abs/2021ApJ...906..131M} {906, 131}

\bibitem[\protect\citeauthoryear{{M{\"u}cke} \& {Protheroe}}{{M{\"u}cke} \&
  {Protheroe}}{2001}]{2001APh....15..121M}
{M{\"u}cke} A.,  {Protheroe} R.~J.,  2001, \mn@doi [Astroparticle Physics]
  {10.1016/S0927-6505(00)00141-9}, \href
  {https://ui.adsabs.harvard.edu/abs/2001APh....15..121M} {15, 121}

\bibitem[\protect\citeauthoryear{{Murase} \& {Stecker}}{{Murase} \&
  {Stecker}}{2022}]{MuraseStecker2022}
{Murase} K.,  {Stecker} F.~W.,  2022, arXiv e-prints, \href
  {https://ui.adsabs.harvard.edu/abs/2022arXiv220203381M} {p. arXiv:2202.03381}

\bibitem[\protect\citeauthoryear{{Narayan} \& {Yi}}{{Narayan} \&
  {Yi}}{1994}]{Narayan1994}
{Narayan} R.,  {Yi} I.,  1994, \mn@doi [\apjl] {10.1086/187381}, \href
  {https://ui.adsabs.harvard.edu/abs/1994ApJ...428L..13N} {428, L13}

\bibitem[\protect\citeauthoryear{{Nilsson}, {Pursimo}, {Villforth}, {Lindfors},
  {Takalo}  \& {Sillanp{\"a}{\"a}}}{{Nilsson}
  et~al.}{2012}]{2012A&A...547A...1N}
{Nilsson} K.,  {Pursimo} T.,  {Villforth} C.,  {Lindfors} E.,  {Takalo} L.~O.,
   {Sillanp{\"a}{\"a}} A.,  2012, \mn@doi [\aap] {10.1051/0004-6361/201219848},
  \href {https://ui.adsabs.harvard.edu/abs/2012A&A...547A...1N} {547, A1}

\bibitem[\protect\citeauthoryear{{Padovani} \& {Giommi}}{{Padovani} \&
  {Giommi}}{1995}]{PadovaniGiommi1995}
{Padovani} P.,  {Giommi} P.,  1995, \mn@doi [\apj] {10.1086/175631}, \href
  {https://ui.adsabs.harvard.edu/abs/1995ApJ...444..567P} {444, 567}

\bibitem[\protect\citeauthoryear{{Padovani} et~al.,}{{Padovani}
  et~al.}{2017}]{AGNReview}
{Padovani} P.,  et~al., 2017, \mn@doi [\aapr] {10.1007/s00159-017-0102-9},
  \href {http://adsabs.harvard.edu/abs/2017A%26ARv..25....2P} {25, 2}

\bibitem[\protect\citeauthoryear{{Padovani}, {Giommi}, {Resconi}, {Glauch},
  {Arsioli}, {Sahakyan}  \& {Huber}}{{Padovani} et~al.}{2018}]{Dissecting}
{Padovani} P.,  {Giommi} P.,  {Resconi} E.,  {Glauch} T.,  {Arsioli} B.,
  {Sahakyan} N.,   {Huber} M.,  2018, \mn@doi [\mnras] {10.1093/mnras/sty1852},
  \href {http://adsabs.harvard.edu/abs/2018MNRAS.480..192P} {480, 192}

\bibitem[\protect\citeauthoryear{{Padovani}, {Oikonomou}, {Petropoulou},
  {Giommi}  \& {Resconi}}{{Padovani} et~al.}{2019}]{Padovani2019}
{Padovani} P.,  {Oikonomou} F.,  {Petropoulou} M.,  {Giommi} P.,   {Resconi}
  E.,  2019, \mn@doi [\mnras] {10.1093/mnrasl/slz011}, \href
  {https://ui.adsabs.harvard.edu/abs/2019MNRAS.484L.104P} {484, L104}

\bibitem[\protect\citeauthoryear{{Padovani} et~al.,}{{Padovani}
  et~al.}{2022a}]{Padovani2022a}
{Padovani} P.,  et~al., 2022a, \mn@doi [\mnras] {10.1093/mnras/stab3630}, \href
  {https://ui.adsabs.harvard.edu/abs/2022MNRAS.510.2671P} {510, 2671}

\bibitem[\protect\citeauthoryear{{Padovani}, {Boccardi}, {Falomo}  \&
  {Giommi}}{{Padovani} et~al.}{2022b}]{Padovani2022b}
{Padovani} P.,  {Boccardi} B.,  {Falomo} R.,   {Giommi} P.,  2022b, \mn@doi
  [\mnras] {10.1093/mnras/stac376}, \href
  {https://ui.adsabs.harvard.edu/abs/2022MNRAS.511.4697P} {511, 4697}

\bibitem[\protect\citeauthoryear{{Perlman} \& {Stocke}}{{Perlman} \&
  {Stocke}}{1994}]{Perlman1994}
{Perlman} E.~S.,  {Stocke} J.~T.,  1994, \mn@doi [\aj] {10.1086/117044}, \href
  {https://ui.adsabs.harvard.edu/abs/1994AJ....108...56P} {108, 56}

\bibitem[\protect\citeauthoryear{{Petkov}, {Novoseltsev}, {Novoseltseva}  \&
  {Baksan Underground Scintillation Telescope Group}}{{Petkov}
  et~al.}{2021}]{ATel15143}
{Petkov} V.~B.,  {Novoseltsev} Y.~F.,  {Novoseltseva} R.~V.,   {Baksan
  Underground Scintillation Telescope Group} 2021, The Astronomer's Telegram,
  \href {https://ui.adsabs.harvard.edu/abs/2021ATel15143....1P} {15143, 1}

\bibitem[\protect\citeauthoryear{{Petropoulou}}{{Petropoulou}}{2014}]{2014A&A...571A..83P}
{Petropoulou} M.,  2014, \mn@doi [\aap] {10.1051/0004-6361/201424603}, \href
  {https://ui.adsabs.harvard.edu/abs/2014A&A...571A..83P} {571, A83}

\bibitem[\protect\citeauthoryear{{Petropoulou}, {Dimitrakoudis}, {Padovani},
  {Mastichiadis}  \& {Resconi}}{{Petropoulou}
  et~al.}{2015}]{2015MNRAS.448.2412P}
{Petropoulou} M.,  {Dimitrakoudis} S.,  {Padovani} P.,  {Mastichiadis} A.,
  {Resconi} E.,  2015, \mn@doi [\mnras] {10.1093/mnras/stv179}, \href
  {https://ui.adsabs.harvard.edu/abs/2015MNRAS.448.2412P} {448, 2412}

\bibitem[\protect\citeauthoryear{{Petropoulou}, {Oikonomou}, {Mastichiadis},
  {Murase}, {Padovani}, {Vasilopoulos}  \& {Giommi}}{{Petropoulou}
  et~al.}{2020}]{2020ApJ...899..113P}
{Petropoulou} M.,  {Oikonomou} F.,  {Mastichiadis} A.,  {Murase} K.,
  {Padovani} P.,  {Vasilopoulos} G.,   {Giommi} P.,  2020, \mn@doi [\apj]
  {10.3847/1538-4357/aba8a0}, \href
  {https://ui.adsabs.harvard.edu/abs/2020ApJ...899..113P} {899, 113}

\bibitem[\protect\citeauthoryear{{Plavin}, {Kovalev}, {Kovalev}  \&
  {Troitsky}}{{Plavin} et~al.}{2020}]{Plavin2020}
{Plavin} A.,  {Kovalev} Y.~Y.,  {Kovalev} Y.~A.,   {Troitsky} S.,  2020,
  \mn@doi [\apj] {10.3847/1538-4357/ab86bd}, \href
  {https://ui.adsabs.harvard.edu/abs/2020ApJ...894..101P} {894, 101}

\bibitem[\protect\citeauthoryear{{Poole} et~al.,}{{Poole}
  et~al.}{2008}]{2008MNRAS.383..627P}
{Poole} T.~S.,  et~al., 2008, \mn@doi [\mnras]
  {10.1111/j.1365-2966.2007.12563.x}, \href
  {https://ui.adsabs.harvard.edu/abs/2008MNRAS.383..627P} {383, 627}

\bibitem[\protect\citeauthoryear{{Predehl} et~al.,}{{Predehl}
  et~al.}{2021}]{2021A&A...647A...1P}
{Predehl} P.,  et~al., 2021, \mn@doi [\aap] {10.1051/0004-6361/202039313},
  \href {https://ui.adsabs.harvard.edu/abs/2021A&A...647A...1P} {647, A1}

\bibitem[\protect\citeauthoryear{{Rector} \& {Stocke}}{{Rector} \&
  {Stocke}}{2001}]{Rector2001}
{Rector} T.~A.,  {Stocke} J.~T.,  2001, \mn@doi [\aj] {10.1086/321179}, \href
  {https://ui.adsabs.harvard.edu/abs/2001AJ....122..565R} {122, 565}

\bibitem[\protect\citeauthoryear{{Rodrigues}, {Garrappa}, {Gao}, {Paliya},
  {Franckowiak}  \& {Winter}}{{Rodrigues} et~al.}{2021}]{2021ApJ...912...54R}
{Rodrigues} X.,  {Garrappa} S.,  {Gao} S.,  {Paliya} V.~S.,  {Franckowiak} A.,
   {Winter} W.,  2021, \mn@doi [\apj] {10.3847/1538-4357/abe87b}, \href
  {https://ui.adsabs.harvard.edu/abs/2021ApJ...912...54R} {912, 54}

\bibitem[\protect\citeauthoryear{{Sahakyan}}{{Sahakyan}}{2018}]{2018ApJ...866..109S}
{Sahakyan} N.,  2018, \mn@doi [\apj] {10.3847/1538-4357/aadade}, \href
  {https://ui.adsabs.harvard.edu/abs/2018ApJ...866..109S} {866, 109}

\bibitem[\protect\citeauthoryear{{Savchenko}, {Larionova}  \&
  {Grisnina}}{{Savchenko} et~al.}{2021}]{ATel15021}
{Savchenko} S.~S.,  {Larionova} E.~G.,   {Grisnina} T.~S.,  2021, The
  Astronomer's Telegram, \href
  {https://ui.adsabs.harvard.edu/abs/2021ATel15021....1S} {15021, 1}

\bibitem[\protect\citeauthoryear{{Stathopoulos}, {Petropoulou}, {Giommi},
  {Vasilopoulos}, {Padovani}  \& {Mastichiadis}}{{Stathopoulos}
  et~al.}{2022}]{2022MNRAS.510.4063S}
{Stathopoulos} S.~I.,  {Petropoulou} M.,  {Giommi} P.,  {Vasilopoulos} G.,
  {Padovani} P.,   {Mastichiadis} A.,  2022, \mn@doi [\mnras]
  {10.1093/mnras/stab3404}, \href
  {https://ui.adsabs.harvard.edu/abs/2022MNRAS.510.4063S} {510, 4063}

\bibitem[\protect\citeauthoryear{{Stecker}, {Done}, {Salamon}  \&
  {Sommers}}{{Stecker} et~al.}{1991}]{Stecker1991}
{Stecker} F.~W.,  {Done} C.,  {Salamon} M.~H.,   {Sommers} P.,  1991, \mn@doi
  [\prl] {10.1103/PhysRevLett.66.2697}, \href
  {https://ui.adsabs.harvard.edu/abs/1991PhRvL..66.2697S} {66, 2697}

\bibitem[\protect\citeauthoryear{{Tavecchio}}{{Tavecchio}}{2006}]{2006tmgm.meet..512T}
{Tavecchio} F.,  2006, in The Tenth Marcel Grossmann Meeting. On recent
  developments in theoretical and experimental general relativity, gravitation
  and relativistic field theories. p.~512 (\mn@eprint {arXiv}
  {astro-ph/0401590}), \mn@doi{10.1142/9789812704030\_0031}

\bibitem[\protect\citeauthoryear{Urry \& Padovani}{Urry \&
  Padovani}{1995}]{Urry1995}
Urry C.~M.,  Padovani P.,  1995, PASP, 107, 803

\bibitem[\protect\citeauthoryear{{Voges} et~al.,}{{Voges} et~al.}{1999}]{RASS}
{Voges} W.,  et~al., 1999, \aap, \href
  {https://ui.adsabs.harvard.edu/abs/1999A&A...349..389V} {349, 389}

\bibitem[\protect\citeauthoryear{{White}, {Giommi}  \& {Angelini}}{{White}
  et~al.}{2000}]{WGA}
{White} N.~E.,  {Giommi} P.,   {Angelini} L.,  2000, VizieR Online Data
  Catalog, \href {https://ui.adsabs.harvard.edu/abs/2000yCat.9031....0W} {p.
  IX/31}

\bibitem[\protect\citeauthoryear{{Xue}, {Liu}, {Petropoulou}, {Oikonomou},
  {Wang}, {Wang}  \& {Wang}}{{Xue} et~al.}{2019}]{2019ApJ...886...23X}
{Xue} R.,  {Liu} R.-Y.,  {Petropoulou} M.,  {Oikonomou} F.,  {Wang} Z.-R.,
  {Wang} K.,   {Wang} X.-Y.,  2019, \mn@doi [\apj] {10.3847/1538-4357/ab4b44},
  \href {https://ui.adsabs.harvard.edu/abs/2019ApJ...886...23X} {886, 23}

\makeatother
\end{thebibliography}








\bsp	
\label{lastpage}
\end{document}